# The Bogdanov Map: Bifurcations, Mode Locking, and Chaos in a Dissipative System

David K. Arrowsmith,
*School of Mathematical Sciences*
*Queen Mary and Westfield College*
*University of London*
*Mile End Road*
*London E1 4NS, UK*

Julyan H. E. Cartwright,
*Departament de Física*
*Universitat de les Illes Balears*
*07071 Palma de Mallorca, Spain*

Alexis N. Lansbury,
*Department of Physics*
*Brunel, The University of West London*
*Uxbridge, Middlesex UB8 3PH, UK*

& Colin M. Place.[*]

We investigate the bifurcations and basins of attraction in the Bogdanov map, a planar quadratic map which is conjugate to the Hénon area-preserving map in its conservative limit. It undergoes a Hopf bifurcation as dissipation is added, and exhibits the panoply of mode locking, Arnold tongues, and chaos as an invariant circle grows out, finally to be destroyed in the homoclinic tangency of the manifolds of a remote saddle point. The Bogdanov map is the Euler map of a two-dimensional system of ordinary differential equations first considered by Bogdanov and Arnold in their study of the versal unfolding of the double-zero-eigenvalue singularity, and equivalently of a vector field invariant under rotation of the plane by an angle $2\pi$. It is a useful system in which to observe the effect of dissipative perturbations on Hamiltonian structure. In addition, we argue that the Bogdanov map provides a good approximation to the dynamics of the Poincaré maps of periodically forced oscillators.

[*]*Formerly: Department of Mathematics, Westfield College, University of London, UK.*

# 1. Introduction

IT is now well known that the study of systems of ordinary differential equations, which commonly arise in dynamical systems investigated in many fields of science, can be aided by utilizing the surface-of-section technique of Poincaré maps. The *Poincaré* or *return* map of a system of ordinary differential equations reduces the dimension of the problem, replacing an $n$-dimensional set of ordinary differential equations with an $(n-1)$-dimensional set of difference equations. The dynamical behaviour of this map can then be investigated much more easily than that of the original system, in the knowledge that the limit sets of the Poincaré map are simply related to the limit sets of the flow. Since it is usually not possible to obtain analytical expressions for Poincaré maps, it is of great interest to construct maps which show similar behaviour, and which are easily computable.

Many problems can be reduced to the study of the dynamics of a system of periodically forced coupled nonlinear oscillators. In this case it is very unlikely that the Poincaré map will be analytically obtainable[1]. A common procedure in this situation is to use the method of averaging on the original system to obtain a set of autonomous differential equations [Hale, 1969; Bogoliubov & Mitropolsky, 1985; Sanders & Verhulst, 1985] by averaging over time. The averaging theorem then guarantees that certain structurally-stable features of the original system will appear as similar features of the averaged system. For instance, a periodic orbit in the original system is represented as a fixed point of the same stability in the averaged system, and bifurcations of the periodic orbit will be reflected in bifurcations of the fixed point in the averaged equations at nearby parameter values. Thus we can see that the time-1 map of the averaged equations is a good approximation to the Poincaré map of the forced oscillator. However, the non-structurally-stable behaviour, such as saddle connections, that is commonly found in averaged equations of Hamiltonian systems [Hayashi, 1985], is not generally seen in the original system, which instead has transverse homoclinic points and Smale horseshoes [Holmes & Holmes, 1981].

Non-structurally-stable behaviour is also not generally preserved in discretizing a system of ordinary differential equations; saddle connections are nongeneric in maps, and discretization acts to restore genericity [Cartwright & Piro, 1992] so that a saddle connection is generically replaced by a homoclinic tangle. Thus we can suggest a connection between the Poincaré map of a forced oscillator, and the discretized version of its averaged equations. If we discretize using the Euler method, the result of this procedure, the *Euler map* of the averaged equations, will be a map with the same fixed points as the averaged equations. We anticipate that the structurally-stable features in the averaged equations will appear in the Euler map, but also that the restoration of genericity forced by the discretization will cause the dynamics of the Euler map to be closer to the dynamics of the Poincaré map of the original forced oscillator than are the dynamics of the averaged equations.

---

[1]One case in which a Poincaré map is analytically obtainable is given by González & Piro [1983].

It was with these ideas in mind that a discrete version of a system of ordinary differential equations first considered by Bogdanov was presented by Arrowsmith & Place [1990b]. The original continuous system, the Bogdanov vector field, is a good example of the behaviour typically to be expected in averaged equations. It has some generality as the archetype of an ordinary differential equation with a sink that undergoes a Hopf bifurcation, the resulting attracting invariant circle growing out to force a saddle-connection bifurcation at a remote saddle in which the circle is destroyed. In this study we analyze in detail the dynamical behaviour of the resulting discrete system, the Bogdanov map.

Our main conclusion is that the Bogdanov map provides a good model for the Poincaré maps of periodically forced oscillators. It has a Hopf bifurcation, like the vector field, but in the map the dynamics on the invariant circle is not trivial. The invariant circle experiences mode locking with the infinity of Arnold tongues formed in the Hopf bifurcation. The saddle connection is replaced by a homoclinic tangle, and the destruction of the invariant circle is consequently far more complicated. We shall analyze in detail the dynamics of the Arnold tongues and we shall show how they are related to the resonances observed in the area-preserving limit. We believe that the Bogdanov map is a useful example of the effect of dissipative perturbations on Hamiltonian structure.

## 2. The Bogdanov Vector Field

A VECTOR FIELD with a fixed point at the origin with the linear part

$$\begin{array}{rcl} \dot{x} & = & y, \\ \dot{y} & = & 0 \end{array} \quad (1)$$

has two degrees of degeneracy; both its determinant and its trace are zero. Generically, we have a codimension-two fixed point at the origin, called a Bogdanov–Takens, cusp[2], or double-zero-eigenvalue singularity, which has the nonzero Jordan canonical form

$$\begin{pmatrix} 0 & 1 \\ 0 & 0 \end{pmatrix}. \quad (2)$$

The normal form for the singularity can be written

$$\begin{array}{rcl} \dot{x} & = & y + ax^2, \\ \dot{y} & = & bx^2, \end{array} \quad (3)$$

---

[2]There is some confusion in the literature here; some authors, for instance Guckenheimer & Holmes [1983], use *cusp* to refer to a different codimension-two fixed point which has one zero eigenvalue, plus a higher-order degeneracy in the normal form.

with $a \neq 0$ and $b \neq 0$. A two-parameter versal unfolding can be given [Arnold, 1972; Takens, 1974; Bogdanov, 1975; Bogdanov, 1976b]

$$\begin{aligned} \dot{x} &= y + v_2 x + ax^2, \\ \dot{y} &= v_1 + bx^2. \end{aligned} \qquad (4)$$

The unfolding given above is not unique, and in fact a different versal unfolding is given in some of the references above, but *versal* means universal, and a versal *unfolding* or *deformation*, such as Eq.(4), contains all possible qualitative dynamical behaviour that can occur near Eq.(3). Making a complete circuit about the origin in $(v_1, v_2)$ space, there exist in sequence four bifurcation curves originating at the origin of saddle–node, Hopf, saddle-connection, and saddle–node type respectively. At parameter values between the two saddle–node curves, there exist no fixed points in the vector field; elsewhere there are two fixed points. As well as being a versal unfolding of the cusp singularity, the vector field above is also and equivalently a versal unfolding of a vector field invariant under rotation of the plane by $2\pi/q$ with $q = 1$. Apart from the original references given above, the system is considered in textbooks by Arnold [1983], Guckenheimer & Holmes [1983], Jackson [1989], Arrowsmith & Place [1990b], and Wiggins [1990], amongst other places.

The system of Eq.(4) allows the fixed points to move relative to each other and eventually coalesce at a saddle–node bifurcation. If we restrict our attention to the region away from the saddle–node bifurcations where there are two fixed points, we arrive at the Hamiltonian system of ordinary differential equations considered by Bogdanov [1976a]

$$\begin{aligned} \dot{x} &= y, \\ \dot{y} &= x(x-1), \end{aligned} \qquad (5)$$

and its two-parameter versal unfolding (which is dissipative)

$$\begin{aligned} \dot{x} &= y, \\ \dot{y} &= u_1 y + x(x-1) + u_2 xy Q(x, u_1, u_2) + u_2^2 y^2 \Phi(x, y, u_1, u_2), \end{aligned} \qquad (6)$$

where $Q$ and $\Phi$ are smooth ($C^\infty$) and $Q(x, 0) \equiv 0$. The phase portrait of Eq.(5) is shown in Fig. 1.

Bogdanov described the bifurcational structure of Eq.(6), which we term the Bogdanov vector field. There are two fixed points of the Bogdanov vector field, at $(x, y) = (0, 0)$ and $(1, 0)$. That at $(1, 0)$ is a saddle. The other, at the origin, is a sink for $u_1 < 0$ and a source for $u_1 > 0$. In between, at $u_1 = 0$, there is a Hopf bifurcation which is subcritical for $u_2 > 0$ and supercritical for $u_2 < 0$. Bogdanov proved that there is at most one limit cycle in the system, and that it is destroyed on the line $u_2 = -7u_1 + O(u_1^2)$ by a saddle-connection bifurcation, otherwise known as a homoclinic bifurcation, or blue-sky catastrophe (since the limit cycle disappears off into the blue). The bifurcation diagram and examples of phase portraits of the Bogdanov vector field are shown in Fig. 2.

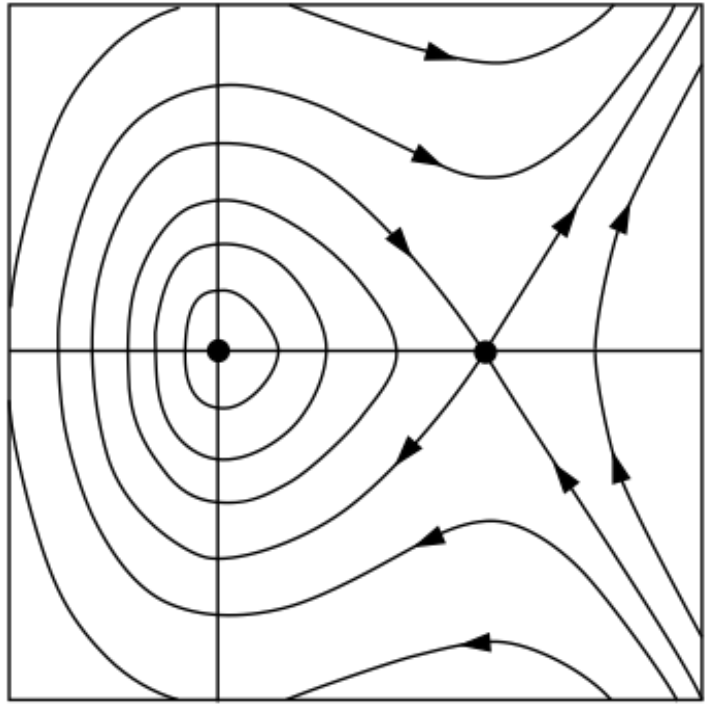

Figure 1: Phase portrait of Eq.(5), which represents a particle in a cubic potential well. The Hamiltonian is $H(x,y) = x^2/2 + y^2/2 - x^3/3$. There is a centre at $(x,y) = (0,0)$, and a saddle at $(1,0)$. The system has a separatrix loop for $H = 1/6$.

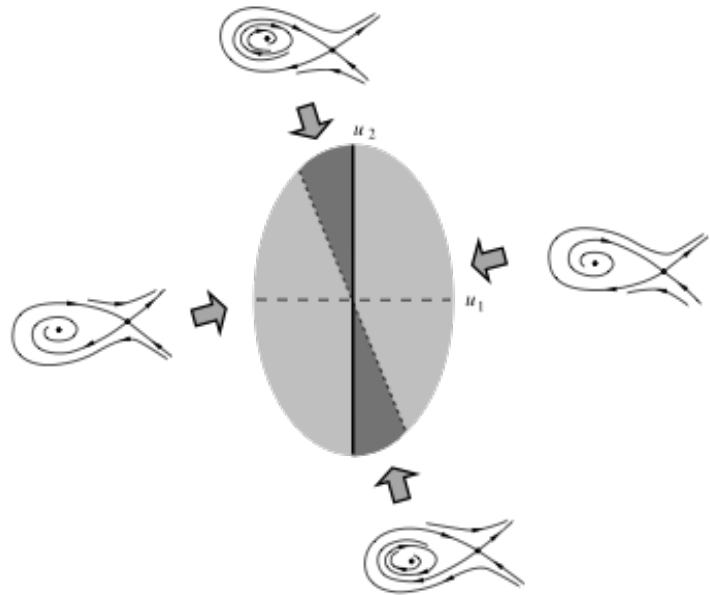

Figure 2: The bifurcation diagram and examples of phase portraits in the Bogdanov vector field, Eq.(6). There is a Hopf bifurcation on the abscissa (solid line), where $u_1 = 0$, and a saddle-connection bifurcation on the slanted dotted line $u_2 = -7u_1 + O(u_1^2)$. Thus there is a limit cycle present in the system only in the sectors shaded darker grey.

The Bogdanov vector field then gives the generic behaviour by which a supercritical Hopf bifurcation creates a stable limit cycle which grows out to be destroyed at a saddle connection. This is a very common component of the behaviour of the averaged equations of a forced oscillator, but saddle connections are not generic in forced oscillators. To obtain behaviour more like that of a forced oscillator, we turn to discretization.

## 3. Derivation of the Bogdanov Map

IF we discretize the Bogdanov vector field of Eq.(6), by applying the backward Euler discretization method to the first equation and the forward Euler method to the second, both with step length $h$, we obtain the Euler map

$$
\begin{aligned}
x' &= x + hy', \\
y' &= y + hu_1 y + hx(x-1) \\
&\quad + hu_2 xy Q(x, u_1, u_2) + hu_2^2 y^2 \Phi(x, y, u_1, u_2).
\end{aligned}
\tag{7}
$$

Set $Q(x, u_1, u_2) = 1$ and $\Phi(x, y, u_1, u_2) = 0$, and let $hy = \tilde{y}$, with $u_1 = \varepsilon/h$, $u_2 = \mu/h$, and $h^2 = k$, where $\varepsilon, \mu, k \in \mathbb{R}$, and (dropping the tilde from $\tilde{y}$) we arrive at the Bogdanov map

$$
\begin{aligned}
x' &= x + y', \\
y' &= y + \varepsilon y + kx(x-1) + \mu xy.
\end{aligned}
\tag{8}
$$

Thus $\varepsilon$ and $\mu$ in the Bogdanov map correspond to $u_1$ and $u_2$ respectively in the Bogdanov vector field. The extra parameter in the map, $k$, plays the role of a step length in the discretization, so we might expect that for small $k$ the behaviour of the map should approximate to that of the vector field. An advantage of the discretization used is that we retain the Hamiltonian nature of Eq.(5) when $\varepsilon = \mu = 0$; in this case, we have a symplectic map. Another advantage of discretizing with the Euler method is that the Bogdanov map, Eq.(8), has only the same two fixed points, at $(x, y) = (0, 0)$ and $(1, 0)$, as the Bogdanov vector field, Eq.(6). It should not be inferred from this that the dynamics of the Euler map are in general the same as those of the differential equations, since it is well-known that the dynamical behaviour of discretized systems can be radically different from that of the original system of ordinary differential equations, even at small step lengths [Cartwright & Piro, 1992]. We emphasize that it is not our aim with the Bogdanov map to emulate the Bogdanov vector field as closely as possible, but rather to allow the perturbation introduced by discretization to restore the generic situation.

The Jacobian of the Bogdanov map is

$$
\begin{pmatrix}
1 + 2kx - k + \mu y & 1 + \varepsilon + \mu x \\
2kx - k + \mu y & 1 + \varepsilon + \mu x
\end{pmatrix}.
\tag{9}
$$

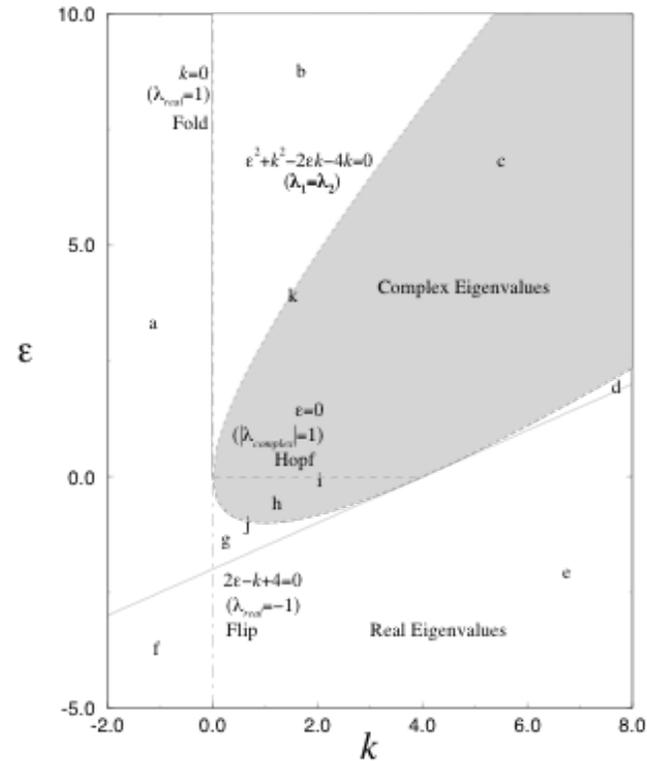

Figure 3: The dependence of the eigenvalues of the fixed point at the origin in the Bogdanov map on the parameters $\varepsilon$ and $k$ for any $\mu$. In regions (a) and (e) the fixed point is a saddle; in (b), (d), and (f), an unstable node; in (c), an unstable focus; in (g), a stable node; and in (h), a stable focus. On the line (i), the fixed point is a centre; on (j), a stable star; and on (k), an unstable star. Regions (g) and (h) together comprise the region of stability in parameter space of the fixed point at the origin.

At the origin this becomes

$$\begin{pmatrix} 1-k & 1+\varepsilon \\ -k & 1+\varepsilon \end{pmatrix} \qquad (10)$$

so we have trace $2+\varepsilon-k$ and determinant $1+\varepsilon$, and the fixed point at the origin is nonhyperbolic on the lines in $(\varepsilon, k)$ space shown in Fig. 3 (there is no $\mu$ dependence). We have complex eigenvalues within the parabola $\varepsilon = k \pm \sqrt{k}$ and real eigenvalues elsewhere. There is a fold bifurcation on the line $k = 0$, a flip (period-doubling) bifurcation on the line $\varepsilon = k/2 - 2$, and a Hopf bifurcation on the line $\varepsilon = 0$ when the eigenvalues are complex, in the interval $0 < k < 4$. We shall show below that we have subcritical Hopf bifurcations at the origin for $\varepsilon = 0$ when $\mu > 0$, and supercritical Hopf bifurcations when $\mu < 0$.

At $(1,0)$ the Jacobian is

$$\begin{pmatrix} 1+k & 1+\varepsilon+\mu \\ k & 1+\varepsilon+\mu \end{pmatrix} \qquad (11)$$

so we have trace $2+\varepsilon+\mu+k$ and determinant $1+\varepsilon+\mu$. This is equivalent to the previous situation if we let $k \to -k$ and $\varepsilon \to \varepsilon+\mu$ in the discussion of the fixed point at the origin. We can mentally reflect the picture in the line $k = 0$, to obtain the switch from $k$ to $-k$, and replace the $\varepsilon$ axis in Fig. 3 with an $\varepsilon + \mu$ axis—this works because $\varepsilon$ and $\mu$ always enter the equations as the pair $\varepsilon + \mu$. This is equivalent to lines in $(\varepsilon, k)$ space becoming surfaces in $(\varepsilon, \mu, k)$ space orientated in the plane $\varepsilon + \mu = 0$.

Thus both fixed points can undergo identical sets of bifurcations in parameter space. We can choose the parameters so that one fixed point experiences all the bifurcations, while the other has none, and we do this by having $k > 0$ so that the fixed point at $(1,0)$ is always a saddle, while the fixed point at the origin can become nonhyperbolic as described above. The choice $k < 0$ would have led to the same thing in reverse; the fixed point at the origin would always be a saddle, while the fixed point at $(1,0)$ could become nonhyperbolic and bifurcate. This is due to there being a symmetry in the map under the transformations $(\bar{x}, \bar{y}) \equiv (1-x, -y)$, which switches the fixed points, and then $(\bar{\varepsilon}, \bar{\mu}, \bar{k}) \equiv (\varepsilon+\mu, -\mu, -k)$, which alters the parameters accordingly. Alternatively, we can note that negative $k$ corresponds to imaginary step lengths in the discretization of the Bogdanov vector field, and argue for $k > 0$ on those grounds.

We mentioned above that the area-preserving nature of the Hamiltonian vector field, Eq.(5), is retained in the discretization. The determinant of the Jacobian of the Bogdanov map is $1+\varepsilon+\mu x$, so if $\varepsilon = \mu = 0$, we have a globally-area-preserving map. We shall look first at the changes that discretization brings to the behaviour of Eq.(5) shown in Fig. 1, by considering the area-preserving Bogdanov map.

# 4. The Area-Preserving Bogdanov Map

THE Bogdanov map is linked closely to the Hénon area-preserving map [Hénon, 1969]

$$\begin{aligned} u' &= u\cos\alpha - (v - u^2)\sin\alpha, \\ v' &= u\sin\alpha + (v - u^2)\cos\alpha. \end{aligned} \qquad (12)$$

Hénon showed that any second-degree area-preserving planar map

$$\begin{aligned} s' &= as + bt + cs^2 + dst + et^2, \\ t' &= fs + gt + hs^2 + ist + jt^2, \end{aligned} \qquad (13)$$

with a centre (elliptic fixed point) at the origin can be reduced to Eq.(12) by a linear change of coordinates. The Bogdanov map with $\varepsilon = \mu = 0$

$$\begin{aligned} x' &= x + y', \\ y' &= y + kx(x - 1), \end{aligned} \qquad (14)$$

and $0 < k < 4$ is such a map. The coordinate transformation between the area-preserving Bogdanov map, Eq.(14), and the Hénon area-preserving map, Eq.(12), is

$$\begin{pmatrix} x \\ y \end{pmatrix} = \frac{1}{4} \begin{pmatrix} 2\sqrt{(4-k)/k} & 0 \\ \sqrt{(4-k)k} & k - 4 \end{pmatrix} \begin{pmatrix} u \\ v \end{pmatrix}, \qquad (15)$$

where in Eq.(12), $\cos\alpha = 1 - k/2$ and $\sin\alpha = \sqrt{(k - k^2/4)}$. In terms of $\alpha$, the transformation becomes

$$\begin{pmatrix} x \\ y \end{pmatrix} = \frac{1}{2} \begin{pmatrix} \operatorname{cosec}\alpha + \cot\alpha & 0 \\ \sin\alpha & -1 - \cos\alpha \end{pmatrix} \begin{pmatrix} u \\ v \end{pmatrix}. \qquad (16)$$

The transformations are nonsingular for $0 < k < 4$, that is, $0 < \alpha < \pi$. If $k < 0$ or $k > 4$, the origin in the area-preserving case is a saddle. Greene *et al.* [1981] have looked at a different quadratic area-preserving map which exhibits the full range of behaviour that can occur when the origin is allowed to become unstable. In the coordinate system of the Hénon area-preserving map, one fixed point remains at the origin, as in the Bogdanov coordinates. The other, however, is no longer fixed at $(1, 0)$; it is now to be found at $(2\tan(\alpha/2), 2\tan^2(\alpha/2))$, and so it coalesces with the fixed point at the origin when $\alpha = 0$.

Note that the Bogdanov map in its general form of nonzero $\varepsilon$ and $\mu$ is not conjugate to Hénon's dissipative map [Hénon, 1976], which he designed as the most general quadratic planar map with constant Jacobian. Although the Bogdanov map is a quadratic planar map, the determinant of the Jacobian matrix in the Bogdanov map is $1 + \varepsilon + \mu x$, which is not a constant.

The Hénon area-preserving map, and thus also the Bogdanov map with $\varepsilon = \mu = 0$, exhibits all the complexity that was first glimpsed by Poincaré [1894] of Hamiltonian dynamical structure in the neighbourhood of a centre. The complexity is addressed in the Poincaré–Birkhoff theorem and the KAM (Kolmogorov–Arnold–Moser) theorem, here in its guise as the Moser twist theorem. Before reminding ourselves of this, it is useful to introduce a concept known as the *rotation number*, or *winding number* of the map. To obtain the rotation number one must write the map in terms of polar coordinates $(r, \theta)$ such that the $\theta$ coordinate is continuous on the real line, that is, it has no jumps. This is known as the *lift* of the map. The rotation number is then defined as

$$\rho(r,\theta) = \lim_{n \to \infty} \frac{\theta_n - \theta_0}{2\pi n}. \tag{17}$$

$\rho$ is the rotation number of the point $(r, \theta)$, when this limit exists. If the limit does not exist, then that point does not have a rotation number. The rotation number is measuring the average rotation per iteration of the map. In an integrable Hamiltonian system trajectories are regular. They lie on smooth curves in phase space known variously as *KAM curves*, *KAM circles*, or *KAM tori*. (The adjective *invariant* can replace *KAM* here, but may also refer to similar structures in dissipative systems, as we shall see below.) The KAM theorem tells us that for sufficiently small nonintegrable perturbations of nonlinear integrable Hamiltonian systems, most of the KAM tori survive. Originally, *sufficiently small* in the proof was *extremely* small, but recent proofs have become much more realistic in this respect. It is the KAM tori with irrational rotation numbers that persist in the perturbed system. They are gradually destroyed as the perturbation is increased, with more irrational KAM tori surviving longer than those nearer to rationals. The most irrational number is the *golden mean*, and it is either the golden-mean KAM torus, or another torus which has a *noble* rotation number that is locally the most irrational, that survives the longest.

Tori with rational rotation numbers, on the other hand, break up immediately. The Poincaré–Birkhoff theorem tells us that they leave behind an even number of periodic points, forming *Birkhoff* periodic orbits. Generically, fixed points and periodic points in area-preserving systems are either centres (elliptic fixed points and periodic points) or saddles (hyperbolic[3] fixed points and periodic points). Centres and saddles alternate on the chains of periodic points in Birkhoff periodic orbits, forming structures known as *island chains*. Around each centre are more KAM tori interspersed with more island chains, each containing centres, around which the structure just described is repeated ad infinitum; the KAM torus and island chain structure exists at all scales. There is still more complexity, this time around the saddles. Generically, the stable and unstable manifolds of a saddle become tangled up, producing *homoclinic* points from the intersection of the stable and unstable manifolds of the same saddle, or, in the island chains, *rotary* homoclinic points from the intersection of the unstable manifold of a point in a periodic saddle with the stable manifold of the next point in the Birkhoff periodic orbit. If there is one homoclinic or rotary homoclinic point then there must be an infinity of them, since their forward and reverse iterates are also homoclinic or rotary homoclinic points. The stable and unstable manifolds must oscillate more and more wildly in phase space between intersections which can be found on any section of the manifolds, that is, they are *dense* on the manifolds. It is this picture that Poincaré [1894] saw in his mind and did not want to attempt to draw.

---

[3]This Hamiltonian terminology is confusing in the context of dissipative systems. Saddles are indeed the only *hyperbolic* fixed points (meaning that the moduli of their eigenvalues is not equal to unity) in Hamiltonian systems, but many hyperbolic fixed points in dissipative system are not saddles.

In the Hénon area-preserving map the amount of perturbation increases away from the origin. So, close in we have almost all KAM curves preserved. Further out, KAM curves become more scarce and island chains predominate. Finally, beyond the last KAM curve, orbits are free to escape to infinity. We give an example in Fig. 4, where we show the the phase portrait of the Bogdanov map for $\varepsilon = 0$, $\mu = 0$, and $k = 1.2$, corresponding to $\alpha = \arccos 0.4$ in the Hénon area-preserving map. We can see that the predictions of the KAM theorem are clearly borne out. Hénon showed that the island chains visible in the phase portraits of the Hénon area-preserving map depend on the choice of $\alpha$ (this corresponds to the choice of $k$ in the Bogdanov map), because $\alpha$ determines the rotation number at the origin. The most prominent island chains in Fig. 4 are of periods 6 and 7.

Hénon [1969] has provided a summary picture for all values of $\alpha$, for initial points on the axis of symmetry in the map. This shows whether the initial point leads to escape to infinity, or to an island chain, or to an invariant circle (see also Vázquez *et al.* [1987]). We reproduce a similar picture in Fig. 5, which shows rotation number in the area-preserving Bogdanov map, translated into Hénon's parameters, $\alpha$ and $w$. We have already introduced $\alpha$. $w$ is the distance from the origin along the axis of symmetry, which is at an angle $\alpha/2$ from the $u$ axis, so that $w = u \cos(\alpha/2)$. The nonintegrable perturbation increases with distance from the origin, and $w$ is a measure of this. In the Bogdanov map parameters, $\alpha = \arccos(1 - k/2)$, and $w = 2x/(\csc(\alpha/2) - \sin(\alpha/2))$. We can clearly see in Fig. 5 the linear increase in rotation number at the origin from 0 to 1/2 as we move from $\alpha = 0$ to $\alpha = \pi$ on the line $w = 0$, and the plateaux of rational rotation numbers almost at right angles to this line which correspond to island chains of different periods. We have plotted rotation number zero at points in parameter space for which we were unable to obtain a rotation number because iterates escaped to infinity.

Hénon [1969] found that the overall shape of Fig. 5 is determined by the positions of the period-1 saddle point and the subharmonic saddles of the island chains in the map. For small $\alpha$, it is the position of the period-1 saddle point on the line $Ow$ that delineates the basin boundary of finite behaviour (meaning anything which does not lie in the basin of infinity[4]). As $\alpha$ tends to zero, the period-1 saddle point moves to the origin, so the basin of finite behaviour shrinks to zero. Bazzani *et al.* [1989] perform further analysis using hyperbolic normal forms to study the stability region. They show that as $\alpha$ increases (corresponding to increasing $k$ in the Bogdanov map), the stable and unstable manifolds of the period-1 saddle oscillate more and more wildly in the homoclinic tangle[5] which has replaced the separatrix loop of Eq.(5), causing the basin of finite behaviour to shrink away from the period-1 saddle point. We show the stable and unstable manifolds of the period-1 saddle for $k = 1.2$ in Fig. 6. This picture is at the same parameter values as Fig. 4. The dynamics of points in the homoclinic tangle formed by the manifolds in the area-preserving Bogdanov map has been investigated by Arrowsmith & Place [1990b], confirming the sensitive dependence on initial conditions in the tangle. Period-3 and higher-period (subharmonic) saddles are influential in determining the boundary at higher $\alpha$. Only odd-period island chains appear on Fig. 5, because even-period island chains do not have their centres on the symmetry line $Ow$.

---

[4]These basins are of course not basins of attraction, since there are no attractors in

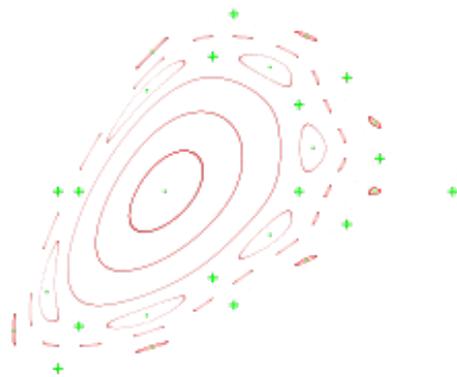

Figure 4: The phase portrait of the Bogdanov map for $\varepsilon = 0$, $\mu = 0$, and $k = 1.2$, corresponding to $\alpha = \arccos 0.4$ in the Hénon area-preserving map. Centres are marked by green dots, saddles by green crosses, and iterates of the map for several initial conditions are shown in red. Island chains of periods 6 and 7 are the most prominent here. The abscissa is $-0.7 < x < 1.3$, and the ordinate is $-1 < y < 1$.

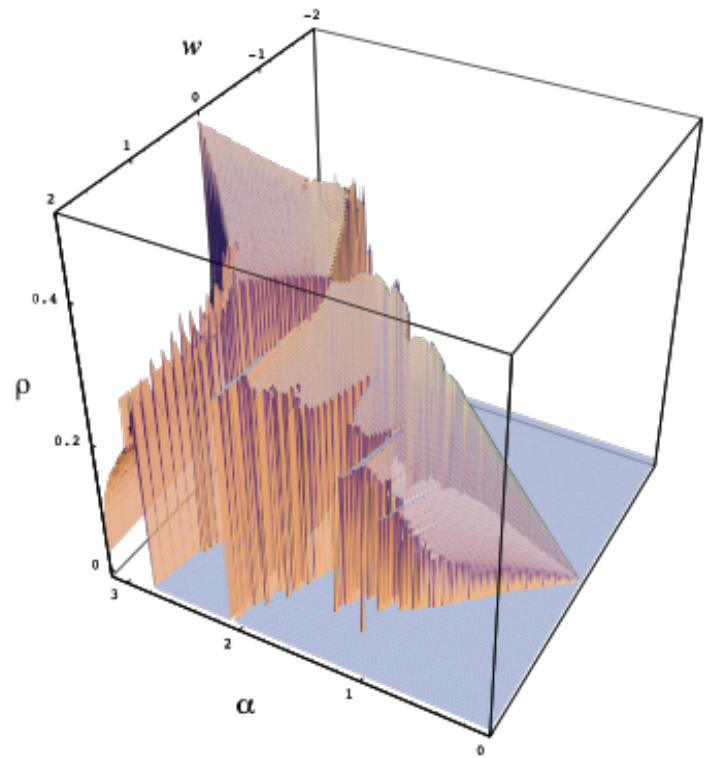

Figure 5: Rotation number in the area-preserving Bogdanov map ($\varepsilon = \mu = 0$) in terms of Hénon's parameters $0 < \alpha < \pi$ and $-2 < w < 2$. Notice the plateaux at rational rotation numbers, and the overall outline of the figure. These may be compared with Fig. 2 in Hénon [1969]. A zero value is plotted at points where iterates escape to infinity.

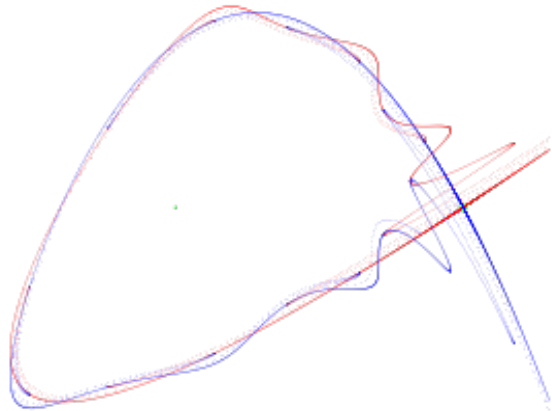

Figure 6: The stable (blue) and unstable (red) manifolds of the period-1 saddle in the Bogdanov map for $\varepsilon = 0$, $\mu = 0$, and $k = 1.2$, corresponding to $\alpha = \arccos 0.4$ in the Hénon area-preserving map, are shown intersecting in a homoclinic tangle. The parameter values are the same as in Fig. 4. The centre at $(0,0)$ is marked by a green dot, and the saddle at $(1,0)$ by a green cross. The abscissa is $-0.7 < x < 1.3$, and the ordinate is $-1 < y < 1$.

Hénon investigated analytically the existence and stability of periodic points up to period 4. He found that there are no period-2 points. Period 3 he found to exist for $\cos\alpha \leqslant 1 - \sqrt{2}$, which corresponds to $k \geqslant 2\sqrt{2}$, as a peculiar island chain with the saddles between the centres and the origin for $-1/2 < \cos\alpha < 1 - \sqrt{2}$ (or $2\sqrt{2} < k < 3$), and with the centres replaced by further saddles for $-1 \leqslant \cos\alpha \leqslant -1/2$ (or $3 \leqslant k \leqslant 4$). Period 4 he found again as a deformed island chain, with its saddles much closer to the origin than its centres. It exists for $\cos\alpha \leqslant 0$ (or $k \geqslant 2$). The centres exist for $-0.10336015\ldots < \cos\alpha < 0$ (or $2 < k < 2.20672030\ldots$), and become another set of saddles for $-1 < \cos\alpha \leqslant -0.10336015\ldots$ (or $2.20672030\ldots \leqslant k < 4$). It is also true of both periods 3 and 4 that all the periodic points tend to the origin as $\alpha \to 2\pi/q$ (or $k \to 2 - 2\cos(2\pi/q)$). It is not possible to solve the equations for periodic points of period greater than 4 analytically, but Hénon's summary picture, our Fig. 5, and further computer investigations show that like period 4, all higher periods are born at the origin at $\alpha = 2\pi p/q$, corresponding to $k = 2 - 2\cos(2\pi p/q)$, and the periodic points move out from the origin as $\alpha$, or $k$, increases. All these higher periods though are normal island chains, not the deformed variety encountered for periods 3 and 4. They begin as island chains of centres and saddles, and at high enough $\alpha$, or $k$, the centres become saddles.

We have discussed the complexity of the dynamics of the Hénon area-preserving map, and we know that the same complexity is found in the area-preserving Bogdanov map. We now turn to the Hopf bifurcation in the dissipative Bogdanov map.

---

conservative systems. Alternative terms for the basin of infinity and the basin of finite behaviour might be the *escaping set* and the *nonescaping set* respectively.

[5]A homoclinic tangle is also known as a *stochastic layer*. There is a lot of interest in describing the motion in the tangle, and in estimating its width as a function of the perturbation.

# 5. The Hopf Bifurcation and Arnold Tongues

THE Hopf bifurcation theorem for maps [Iooss, 1979; Whitley, 1983; Arrowsmith & Place, 1990b; Wiggins, 1990] tells us that if $f_\nu(x, y)$ is a one-parameter family of maps of the plane satisfying:

(a) $f_\nu(0, 0) = (0, 0)$ for $\nu$ near 0;

(b) $Df_\nu(0, 0)$ has two complex eigenvalues $\lambda(\nu)$ and $\bar{\lambda}(\nu)$ for $\nu$ near 0 with $|\lambda(0)| = 1$ and $\lambda = \lambda(0)$ not a $q$th root of unity for $q = 1, 2, 3,$ or 4;

(c) $\left.\dfrac{d|\lambda(\nu)|}{d\nu}\right|_{\nu=0} > 0;$

then we have a Hopf bifurcation occurring at $\nu = 0$. The fixed point at the origin in the Bogdanov map satisfies (a), (b), and (c) at $\varepsilon = 0$ when $\mu \neq 0$ and $0 < k < 4$, so we know that there is a Hopf bifurcation in the Bogdanov map in an analogous situation to the Hopf bifurcation in the Bogdanov vector field. In order to obtain information about the dynamics of the map in the neighbourhood of the Hopf bifurcation, we need to perform a standard normal form calculation, which we show in Appendix A.1.

We obtain the result that the Hopf bifurcation is subcritical for $\mu > 0$ and supercritical for $\mu < 0$ (see Fig. 7). There is a degenerate Hopf bifurcation when $\mu = 0$. The strong resonance cases occur with $q = 3$ at $k = 3$ and $q = 4$ at $k = 2$; $q = 1$ is at $k = 0$ and $q = 2$ has $k = 4$, so these are degenerate points anyway, as we can see from Fig. 3.

When we have an attracting Hopf circle from a supercritical Hopf bifurcation, we have two different possibilities for its dynamics. Either we will have quasiperiodic motion with irrational rotation number and orbits densely covering the circle, or periodic motion with rational rotation number and with saddles and sinks alternating around the circle. In the latter case, $\alpha = \arg \lambda$ is given by $2\pi p/q$.

If we now let $\lambda = (1 + \eta)e^{i(\alpha+\phi)}$, then we obtain local polar coordinates $\eta = |\lambda| - 1$ and $\phi = \arg \lambda - \alpha$ in the neighbourhood of the bifurcation. We use the normal form for a weak resonance (see Appendix A.1)

$$w' = \lambda w + n_{21} w^2 \bar{w} + \cdots + n_{0q-1} \bar{w}^{q-1} + \cdots, \quad (18)$$

take the argument, and manipulate using $w = re^{i\theta}$ and $n_{21} = be^{i\gamma}$ to get

$$\theta' = \theta + \alpha + \arg\left((1+\eta)e^{i\phi} + br^2 e^{i(\gamma-\alpha)} + n_{0q-1} r^{q-2} e^{-i(q\theta+\alpha)}\right). \quad (19)$$

Using the fact that we are on the Hopf circle, so that $r = \sqrt{(-\eta/\mathrm{Re}(\bar{\lambda} n_{21}))}$, we have

$$\theta' = \theta + \alpha + \quad (20)$$
$$\arg\left((1+\eta)e^{i\phi} - \frac{be^{i(\gamma-\alpha)}\eta}{\mathrm{Re}(\bar{\lambda} n_{21})} + \left(-\frac{\eta}{\mathrm{Re}(\bar{\lambda} n_{21})}\right)^{\frac{q-2}{2}} n_{0q-1} e^{-i(q\theta+\alpha)}\right).$$

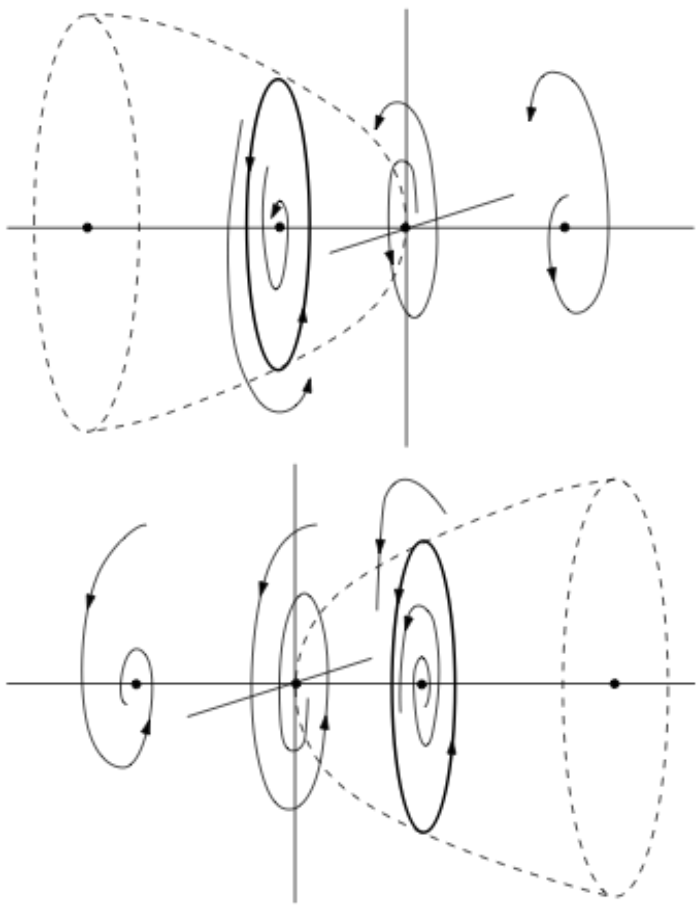

Figure 7: Schematic diagrams of subcritical (upper picture) and supercritical (lower picture) Hopf bifurcations. In the subcritical case an unstable focus changes stability to become a stable focus, and spawns a repelling invariant circle in the process. In the supercritical case it is a stable focus that bifurcates to become an unstable focus, whilst giving birth to an attracting invariant circle.

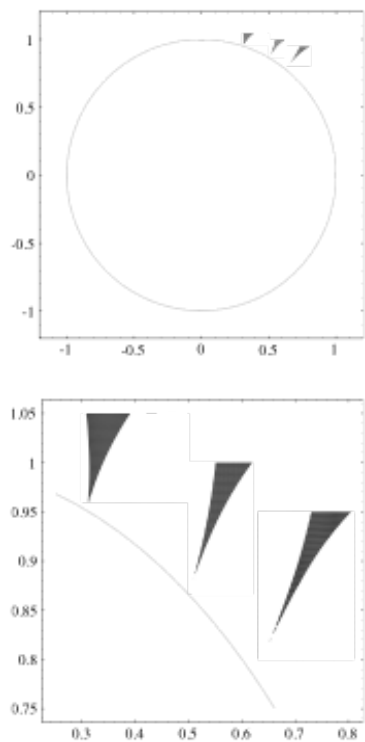

Figure 8: Arnold tongues in $(\text{Re}(\lambda), \text{Im}(\lambda))$ space. The plots show the unit circle from which the tongues emerge, and from left to right in the picture, the $1/5$, $1/6$, and $1/7$ tongues. The lower plot is a closeup view of the upper one. The tongues are solutions of Eq.(23). All the tongues begin at the unit circle, but the numerical algorithm used to plot them finds difficulty in picking them up when they are very thin near the circle.

If we have a periodic orbit of period $q$, then to a first approximation the argument should be zero. Thus the expression inside the parentheses must have zero imaginary part, and so

$$(1+\eta)\sin\phi \quad - \quad \frac{b\sin(\gamma-\alpha)}{\text{Re}(\bar{\lambda}n_{21})}\eta \quad (21)$$
$$+ \quad \left(-\frac{\eta}{\text{Re}(\bar{\lambda}n_{21})}\right)^{\frac{q-2}{2}}|n_{0q-1}|\sin(\arg n_{0q-1} - q\theta - \alpha) = 0.$$

For small $\eta$ and $\phi$ in the neighbourhood of the bifurcation, this leads to the equations, first given by Arnold [1983]

$$\phi = \frac{\text{Im}(\bar{\lambda}n_{21})}{\text{Re}(\bar{\lambda}n_{21})}\eta \pm \frac{|n_{0q-1}|}{|\text{Re}(\bar{\lambda}n_{21})|^{\frac{q-2}{2}}}\eta^{\frac{q-2}{2}}, \quad (22)$$

or equivalently

$$\left|\phi - \frac{\text{Im}(\bar{\lambda}n_{21})}{\text{Re}(\bar{\lambda}n_{21})}\eta\right| \leqslant \frac{|n_{0q-1}|}{|\text{Re}(\bar{\lambda}n_{21})|^{\frac{q-2}{2}}}\eta^{\frac{q-2}{2}}, \quad (23)$$

for the boundaries within which a period-$q$ resonance is found. Plotting Eq.(23) in terms of $\lambda$ leads to the regions shown in Fig. 8. Remembering that the analysis above is valid in the neighbourhood of the bifurcation, which occurs on the unit circle in $\lambda$ space, and that we are considering a Hopf bifurcation that leads to an invariant circle as the eigenvalues move out of the circle, we can interpret Fig. 8 as showing regions variously described as *tongue* or *horn* shaped, with cusps on the unit circle at $\lambda = e^{2i\pi p/q}$. Only tongues outside the circle are valid solutions in this case. In Fig. 8 we show these so-called *Arnold tongues*, otherwise known as *Arnold horns*, *resonance tongues*, or *resonance horns*, for $q = 5$, 6, and 7. Remember that for $q \leqslant 4$ we have the strong resonances whose behaviour is different. Often, Arnold tongues exist in parameter space in a region considerably greater than just that local to the cusp at the unit circle, and they can eventually overlap.

In order to see the Arnold tongues in terms of the Bogdanov map parameters, we can use Eq.(23). We set $\arg \lambda = \arctan\sqrt{4(1+\varepsilon)/(2+\varepsilon-k)^2-1}$ and $|\lambda| = 1+\varepsilon$ to obtain the equation

$$\left|\arctan\sqrt{\frac{4(1+\varepsilon)}{(2+\varepsilon-k)^2}-1} - \frac{2\pi p}{q} - \frac{\text{Im}(\bar{\lambda}n_{21})}{\text{Re}(\bar{\lambda}n_{21})}\varepsilon\right| \leqslant \frac{|n_{0q-1}|}{|\text{Re}(\bar{\lambda}n_{21})|^{\frac{q-2}{2}}}\varepsilon^{\frac{q-2}{2}} \quad (24)$$

for the edges of the tongues near the bifurcation in the Bogdanov map. The same Maple program as given in Appendix A.1, but working to order $q-1$ for a period-$q$ resonance, enables us to obtain $n_{0q-1}$ and $\text{Im}(\bar{\lambda}n_{21})$. Since the number of terms increases very rapidly with the order of the expansion, it is necessary for higher period resonances to be very careful about which intermediate terms are retained; all terms other than those required to obtain $n_{21}$ and $n_{0q-1}$ must be dropped. Even doing this, we find it impossible to obtain results for tongues of period higher than 6, corresponding to $n_{06}$ and above, with a workstation in reasonable time. We plot the beginnings of the period-5 and period-6 tongues in the Bogdanov map in $(\varepsilon, k)$ space in Fig. 9.

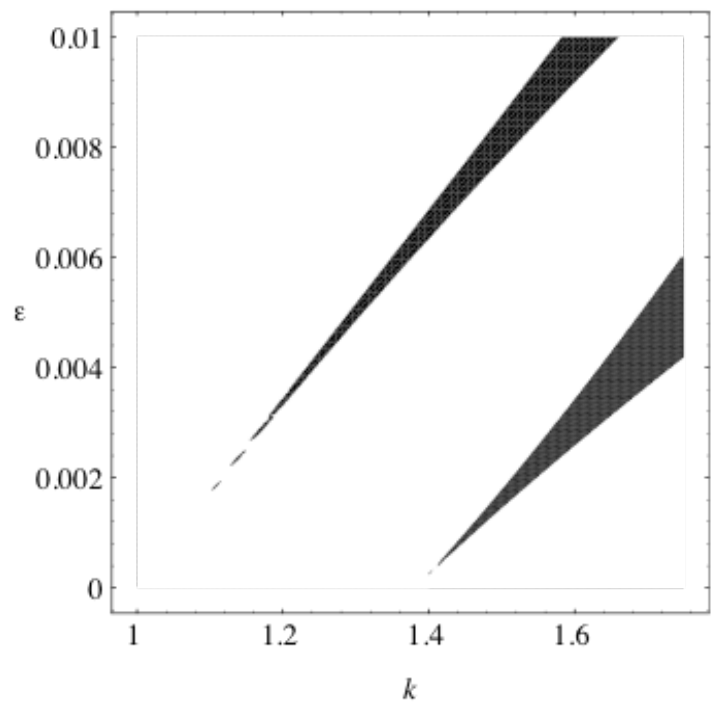

Figure 9: Predicted appearance of Arnold tongues in the Bogdanov map in $(\varepsilon, k)$ space for $\mu = -0.1$. The tongues are solutions of Eq.(24). Shown here are the 1/5 tongue, on the right, and the 1/6 tongue, on the left of the picture. The same comments as in the previous figure apply to the difficulty of picking the tongues up near to their cusps at $\varepsilon = 0$.

We have shown that the Hopf theorem predicts the cuspidal beginnings of the $p/q$ Arnold tongue at $\varepsilon = 0$ when $k = 2 - 2\cos(2\pi p/q)$. To verify this prediction for the Bogdanov map, we can use Newton's method and plot the points in the $(\varepsilon, k)$ parameter space at which a periodic orbit of period $q$ can be found. We display the results of a survey at $\mu = -0.1$ for periods 5 to 10 in Fig. 10. The $1/5$, $1/6$, $1/7$, $1/8$, $1/9$, and $1/10$ tongues are coloured cyan, green, red, yellow, magenta, and blue respectively. Comparison of this picture with that obtained from the Hopf bifurcation theorem, Fig. 9, shows that Eq.(24) correctly reproduces the tongue behaviour near to the bifurcation at $\varepsilon = 0$, so that the angles at which the tongues emerge are predicted. The real tongues however bend back on themselves to cross the line $\varepsilon = 0$ once more at higher values of $k$. This deviation from Fig. 9 is to be expected, since the Hopf theorem only deals with the behaviour local to the bifurcation. The fractal structure that can be seen inside the tongues in Fig. 10 is caused by the method of construction of the picture. The Newton method for finding the periodic orbits was given a fixed initial condition $(x, y) = (0.5, -0.25)$. Most of the time, if a period-$q$ tongue existed at the parameter values investigated, this initial condition was inside its basin of attraction, as is evidenced by the tongues being almost complete. Sometimes however the initial condition was not in the basin of attraction of the tongue, so the Newton method failed to find the tongue at those parameter values. It was this that produced the fractal structure seen inside some of the tongues.

A little more analysis of the circle map in Eq.(21) shows that in an Arnold tongue produced in a supercritical Hopf bifurcation, as stated above, we have two periodic orbits, with saddles and sinks alternating on the invariant circle, so that a saddle has a sink on both sides, and equally a sink has a saddle on both sides. The Hopf bifurcation theorem tells us that if the map is smooth ($C^\infty$), then the invariant circle will also be smooth in the neighbourhood of the bifurcation. This implies that the sinks on an invariant circle near to its birth must be nodes, because the circle would not be differentiable at a focus. At the edges of the tongue, the periodic points come together and coalesce in saddle–node bifurcations. If we look in the neighbourhood of a node, as we move across the tongue from one edge to the other, we see the node born together with a saddle, which then moves away as the parameters change until, at the other side of the tongue, the saddle approaches and coalesces with the next node along in the chain.

There is obviously a link between the island chains in the Hénon area-preserving map, or equivalently the island chains in the area-preserving Bogdanov map, and the Arnold tongues that appear as we add dissipation to produce a Hopf bifurcation. The strong resonances are related to the peculiar island chains that we observe for $q = 3$ and $q = 4$ in the area-preserving case, and the weak resonances to the regular island chains for $q \geqslant 5$. To uncover the mechanism for this link, we must put together the information from both the area-preserving and the dissipative Bogdanov maps to see the whole picture.

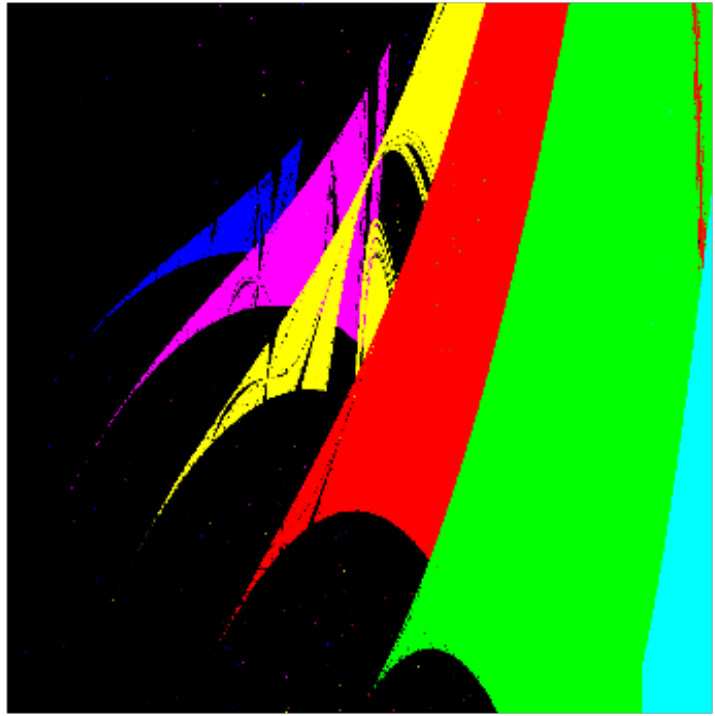

Figure 10: Arnold tongues in the Bogdanov map. The $1/5$, $1/6$, $1/7$, $1/8$, $1/9$, and $1/10$ tongues are shown as computed using Newton's method, in cyan, green, red, yellow, magenta, and blue respectively. The abscissa is $0 < \varepsilon < 0.02$, and the ordinate is $0.5 < k < 1.5$. $\mu$ is set to $-0.1$. The appearance of the $1/5$ and $1/6$ tongues can be compared with the predictions made for their behaviour near to their cusps in Fig. 9.

## 6. The Hopf Circle and Birkhoff Attracting Sets

NOTWITHSTANDING the fact that when $\varepsilon \neq 0$ or $\mu \neq 0$ the Bogdanov map is no longer area-preserving, computer experiments indicate that a large part of the area-preserving structure is retained with weak dissipation[6], but generically the island chains contain foci rather than centres. It is easy to show using the implicit function theorem that the Birkhoff periodic points of the Hénon area-preserving map, and equivalently of the area-preserving Bogdanov map, persist in the dissipative Bogdanov map for $\varepsilon$ and $\mu$ sufficiently close to zero. Let $f_{\varepsilon,\mu}$ be the dissipative Bogdanov map, so that $f_{0,0}$ is the area-preserving Bogdanov map which is conjugate to the Hénon area-preserving map. Let $x = x^*$ be a period-$q$ point of $f_{0,0}$, then $f_{0,0}^q(x^*) = x^*$. Consider the equation

$$g(\varepsilon, \mu, x) = f_{\varepsilon,\mu}^q(x) - x = 0, \qquad (25)$$

and look for a solution of the form $x = x(\varepsilon, \mu)$. We have the solution $x = x^*$ for $\varepsilon = \mu = 0$. Consider now

$$Dg(\varepsilon, \mu, x) = Df_{\varepsilon,\mu}^q(x) - I. \qquad (26)$$

If $x^*$ is a saddle then

$$|Df_{0,0}^q(x^*) - I| \neq 0, \qquad (27)$$

and if $x^*$ is a centre then

$$|Df_{0,0}^q(x^*) - I| = \begin{vmatrix} \cos\alpha - 1 & -\sin\alpha \\ \sin\alpha & \cos\alpha - 1 \end{vmatrix} = 2 - 2\cos\alpha \neq 0 \qquad (28)$$

as long as $\alpha \neq \pm\pi/2$ which corresponds to strong resonance. Now, by the implicit function theorem, there is a smooth function $x = x(\varepsilon, \mu)$ for $(\varepsilon, \mu)$ in a neighbourhood of $(0,0)$ such that $f_{\varepsilon,\mu}^q(x) - x = 0$ in this neighbourhood, as required. So we have shown that the Birkhoff periodic orbits which exist in the Hénon area-preserving map at a particular value of $\alpha$, or in the area-preserving Bogdanov map at the equivalent value of $k$, will continue to exist in the dissipative Bogdanov map at the same value of $k$ for $\varepsilon$ and $\mu$ sufficiently small. We give an example in Fig. 11, where we show the phase portrait for $\varepsilon = 0.01$, $\mu = -0.1$, and $k = 1.2$, corresponding to Fig. 4 with dissipation added. The period-6 and period-7 island chains we saw in Fig. 4 are still present, in only slightly changed positions, but the centres of Fig. 4 are now attracting foci in Fig. 11.

---

[6]The Jacobian determinant is $J = 1 + \varepsilon + \mu x$, which can be either smaller or larger than one. We are using the convention here that if $J \neq 1$ somewhere, then we have dissipation, and if $J < 1$ everywhere then we have *strict* dissipation; the Bogdanov map is then dissipative but not strictly dissipative. Weak dissipation implies that $J \simeq 1$.

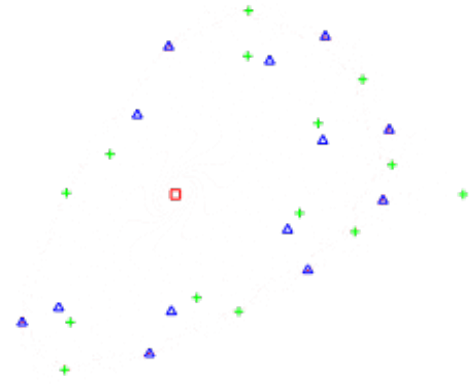

Figure 11: The phase portrait of the Bogdanov map for $\varepsilon = 0.01$, $\mu = -0.1$, and $k = 1.2$, corresponding to Fig. 4 with dissipation added. Sinks are marked by blue triangles, sources by red squares, and saddles by green crosses. Iterates of the map for a single initial condition close to the origin are shown in red. Notice that the period-6 and period-7 island chains are still present in almost the same places as in Fig. 4. The abscissa is $-0.7 < x < 1.3$, and the ordinate is $-1 < y < 1$.

Analysis of the dissipative version of the Bogdanov map compared with the area-preserving case is complicated by the Hopf bifurcation at the origin when $\varepsilon = 0$ and $0 < k < 4$. The Hopf bifurcation theorem predicts the growth of an attracting invariant circle from the origin for a sufficiently small range of positive $\varepsilon$ when $\mu < 0$. In the case of the Bogdanov vector field, we have a continuously growing circle which is destroyed in a saddle connection. The fate of the invariant circle is somewhat different in the map. We find that as the Hopf invariant circle grows out from the origin, it interacts with the Birkhoff periodic orbits in the Hénon island chains in a way which involves changing its structure from being a simple attracting circle to something more complicated—a strange attracting set. To be a *strange* attracting set, it must contain a transverse homoclinic point; a homoclinic point in which the stable and unstable manifolds cross, or intersect tranversely. This situation, where the Birkhoff periodic orbits become part of a more complicated attracting set, is predicted by a theorem of Aronson *et al.* [1983] (see also Arrowsmith & Place [1990a; 1990b]). The theorem gives sufficient conditions for the existence of a Birkhoff attracting set with a nontrivial rotation interval. In order to establish that the invariant circle in the Bogdanov map becomes part of a more complicated Birkhoff attracting set on passing out of resonance with a Birkhoff periodic orbit, one must check that transverse homoclinic points to the periodic orbit occur, and that there is a trapping region[7] containing the invariant circle and the periodic orbit. This is discussed further by Arrowsmith & Place [1990b] for the Bogdanov map, and also by Arrowsmith & Place [1990a] for another map of the plane, the Euler map of a vector field invariant under rotation of the plane by $\pi/2$.

The Hopf circle comes into resonance with an island chain of a particular rotation number in a standard way. First we see rotary homoclinic tangles of the inner branches of the manifolds of the saddles in the island chain. Then the invariant circle comes into resonance with the island chain for an interval of parameter values, forming an attracting set of the saddles and foci, with the foci as the attractor within it. (An attractor is an attracting set which contains a dense orbit, thus making it irreducible, so that the union of two attractors is not an attractor, but an attracting set.) After this, we see rotary homoclinic tangles of the outer branches of the manifolds of the saddles, as the Hopf circle leaves the vicinity of resonance with the Birkhoff periodic orbit. The rotary homoclinic tangles give rise to the strange attracting set and due to them, by the Smale–Birkhoff homoclinic theorem, the map will have embedded horseshoes and infinitely many periodic orbits. The Hopf circle, moving out, approaches an island chain, interacts with it in the manner described above, and then reappears on the other side. We give a schematic diagram of this behaviour in Fig. 12. As the Hopf parameter $\varepsilon$ is increased, the behaviour depicted in Fig. 12 is repeated for the infinity of island chains from the area-preserving case as the Hopf circle moves into and out of resonance. We should emphasize that the interval over which homoclinic tangles occur is very small; we have observed them for intervals in $\varepsilon$ of the order of $10^{-11}$ to $10^{-12}$ at $\mu = -0.1$. A sequence of computer-generated pictures showing this behaviour in the periodic orbit of order $1/6$ in the Bogdanov map was presented by Arrowsmith & Place [1990b]. Similar behaviour was also observed by Arrowsmith & Place [1990a] in the Euler map of the versal unfolding of a vector field invariant under rotation by $2\pi/q$ where $q = 4$.

---

[7]A trapping region is a region that maps into its interior.

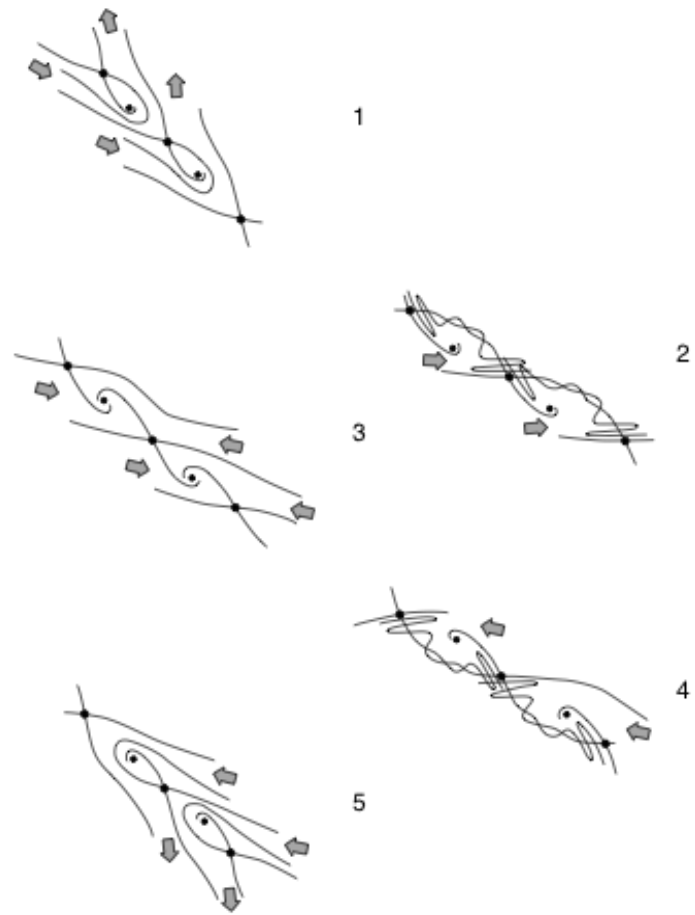

Figure 12: A sequence of schematic diagrams of the invariant circle moving through resonance with an island chain. Larger dots indicate the positions of saddles; smaller dots, sinks. From top to bottom we first see the invariant circle approach an island chain. In the next three diagrams it passes through the island chain. First we see rotary homoclinic tangles of the manifolds of the saddles on one side. In the middle diagram we see the invariant circle in resonance with the periodic orbit. Subsequently we have rotary homoclinic tangles on the other side. Finally, we see the invariant circle emerge on the opposite side of the island chain.

As well as the main Hopf bifurcation, whose extremely complicated dynamics we have just discussed, there are also Hopf bifurcations away from the origin, at the foci of the island chains. These secondary Hopf bifurcations do not necessarily occur at $\varepsilon = 0$ as the primary Hopf bifurcation does, since if we have a periodic orbit of period $q$, its stability is given by

$$\mathrm{Det} = \prod_{i=1}^{q}(1 + \varepsilon + \mu x_i),  \qquad (29)$$

so if some $x_i$s are nonzero, then $\varepsilon$ will not be zero when $\mathrm{Det} = 1$. For example, we show below that a secondary Hopf bifurcation to give attracting invariant circles occurs at $\varepsilon \simeq 0.0032$ for the $1/5$ resonance when $\mu = -0.1$ and $k = 1.44$. These secondary Hopf bifurcations are possible because the Bogdanov map is not strictly dissipative (which would imply that $\mathrm{Det} < 1$ always), so it is possible for the sinks in the primary invariant circle, resulting from a supercritical Hopf bifurcation at the origin, to lose stability and become sources as the primary invariant circle grows out. The secondary invariant circles grow out through secondary island chains from the area-preserving case in exactly the same way as we have described happening for the primary invariant circle and the primary island chains in the map. The same goes for tertiary island chains and tertiary Hopf bifurcations, and so on. This behaviour occurs on all scales, and the same analysis as we performed for the primary invariant circle in the map would show similar resonant behaviour occurring on all the invariant circles. Thus we have an infinite hierarchy of Hamiltonian structure from the area-preserving Bogdanov map, with an infinity of Hopf bifurcations superimposed by the dissipation.

## 7. Resonance and the Devil's Quarry

IN Fig. 13 we have a plot of rotation number at a fixed initial point for different values of $\varepsilon$ and $k$, with $\mu = -0.1$ so that we are in the supercritical Hopf bifurcation regime. This picture, which we term a *devil's quarry*, illustrates how the attracting circle moves into and out of resonance with island chains of different periods as the parameters change. One can see from the devil's quarry that *resonance* (also termed *phase locking*, *mode locking*, or *entrainment*) with an island chain increases in measure as $\varepsilon$ and $k$ increase. Resonances of orders $1/3$, $1/4$, $2/9$, $1/5$, $2/11$, $1/6$, $1/7$, $1/8$, and $1/9$ are most obvious in Fig. 13. In this picture, as in Fig. 5, rotation number zero is plotted where iterates escaped to infinity so that we could not obtain a rotation number.

The term *devil's quarry* comes by analogy with the *devil's staircase* which was originally introduced in circle maps to show the measure of the mode locking at a particular value of the nonlinearity, by plotting rotation number as a function of the forcing frequency. Dissipative systems with competing frequencies show the phenomenon of phase locking, where the system locks into motion which has a rational frequency ratio, first observed and explained by Huygens [1665a] in two pendulum clocks coupled by a common mounting[8]. Phase locking increases with nonlinearity, from no phase locking in the linear regime, to a critical situation where the system is everywhere phase locked. At higher values of nonlinearity, chaotic motion may occur as well as periodic and quasiperiodic motion. In a subcritical[9] circle map, intervals on which the rotation number is constant and rational, where there is a resonance of a particular period, punctuate intervals of monotonically increasing rotation number; the staircase is incomplete. In supercritical circle maps the resonances overlap. This leads to the onset of transient chaos as iterates wander between the overlapping resonances. In a critical circle map, the staircase is complete; the measure of the phase-locked intervals is one, and the rotation number increases in a staircase fashion with steps at each rational rotation number and risers in between. The staircase has an infinity of steps at all scales (so that someone climbing the staircase step by step would never reach the top—a devilish construction indeed!); the size of the step decreases as the period of the associated cycle increases. Between any two steps associated with rational rotation numbers, the largest intermediate step is given by the rational having the smallest denominator in that interval. Such an ordering of rationals is provided by the Farey tree [González & Piro, 1983; Aronson *et al.*, 1983; Cvitanović *et al.*, 1985; Hao, 1989] constructed using the Farey mediant

$$\frac{p}{q} \oplus \frac{p'}{q'} = \frac{p+p'}{q+q'}. \tag{30}$$

To obtain the Farey tree with this rule we start with the two ends of the unit interval written as $0/1$ and $1/1$, to obtain at the first level $1/2$, at the second level $1/3$ and $2/3$, at the third level $1/4$, $2/5$, $3/5$, and $3/4$, and so on.

The devil's staircase becomes the devil's quarry when we look at the picture in terms of the nonlinearity as well as the forcing frequency, so that we have rotation number represented by height in a three-dimensional quarry. In the one-dimensional sine circle map

$$\theta' = \theta + \Omega - \frac{k}{2\pi}\sin 2\pi\theta \bmod 1, \tag{31}$$

---

[8]The 'sympathy of clocks'; the relevant pages from Huygens' notebook are reprinted in [Huygens, 1888–1950].

[9]Unfortunately, the same words are used for two different purposes. The notation *subcritical*, *critical*, or *supercritical*, referring to circle maps, is not related to subcritical and supercritical Hopf bifurcations.

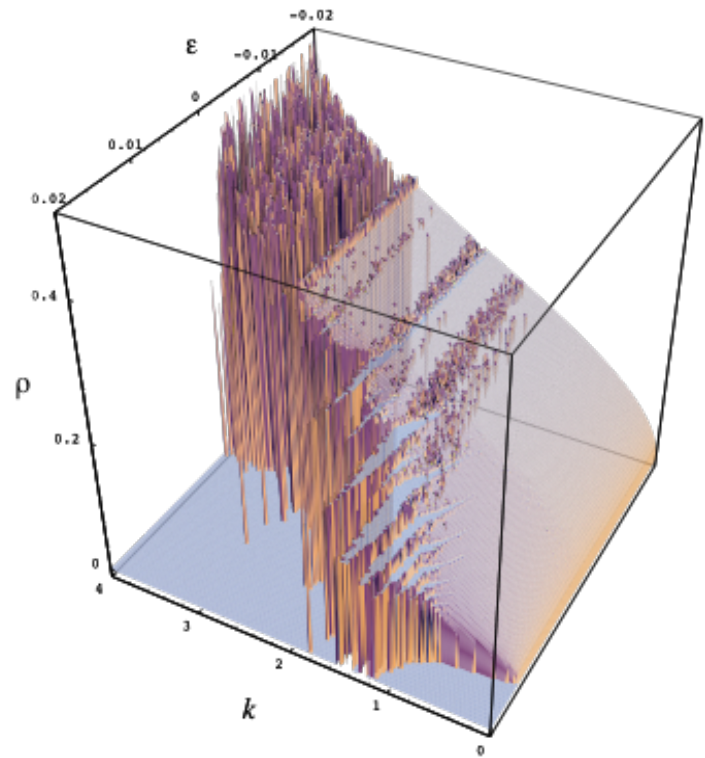

Figure 13: A devil's quarry rotation number plot in the Bogdanov map for $-0.02 < \varepsilon < 0.02$ and $0 < k < 4$ with $\mu = -0.1$. Resonance with rotation numbers $1/3$, $1/4$, $2/9$, $1/5$, $2/11$, $1/6$, $1/7$, $1/8$, and $1/9$ is most obvious in this plot. A zero value is plotted at points where iterates escape to infinity.

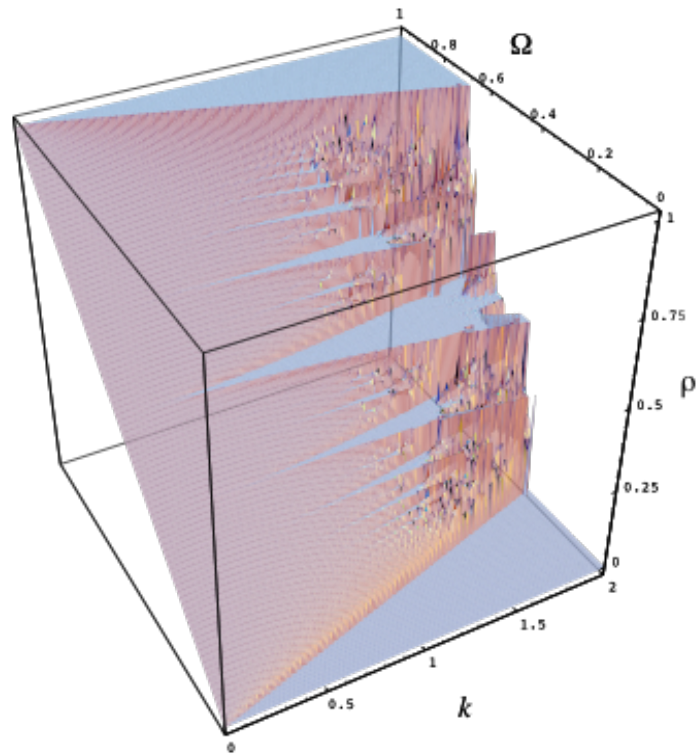

Figure 14: A devil's quarry rotation number plot in the sine circle map for $0 < \Omega < 1$ and $0 < k < 2$. Notice the contrast between the order in the subcritical region $k < 1$, and the chaos of the supercritical region $k > 1$. Note also the accumulation point of period doubling to chaos that is visible in the larger tongues.

the critical case is a straight line across the quarry at $k = 1$, dividing the subcritical from the supercritical behaviour. We show the devil's quarry for the sine circle map in Fig. 14. We can see that the Farey tree provides the local ordering of the step widths as we described above. Chaos in the supercritical circle map is reflected in the devil's quarry as the rough rockface for $k > 1$. Inside the larger resonances, the accumulation point of period doubling leading to chaos is also obvious.

In the Bogdanov map the picture is not quite as simple as in the circle maps, since all parameters add nonlinearity, and because there is no simple relationship between the parameters and the critical line. Bohr *et al.* [1984] found for the dissipative standard map that a critical line on which the staircase is complete does exist and is a smooth curve on the high-order phase-locked intervals for that system. We do not yet know whether this is also the case in the Bogdanov map, or whether instead the critical curve is a fractal. We hope to obtain further results on this point, which will be published elsewhere. We show a more closeup view of the devil's quarry in the Bogdanov map in Fig. 15. We can see much more clearly than in Fig. 13 the structure of the mode locking in parameter space. Resonances with island chains of orders 2/11, 1/6, 1/7, 1/8, 1/9, 1/10, 1/11, 1/12, and 1/13 are the most prominent in Fig. 15. Resonance overlap is clearly visible at the edges of the plateaux, giving the effect of columns and crevices in the quarry. A closer look at these regions shows an infinity of smaller and smaller plateaux from resonances with higher and higher periods, as predicted by the Farey sequence of rotation numbers, accumulating on the boundary of every resonance.

Apart from rotation number, another characteristic of a dynamical system useful for obtaining information about resonance is its Lyapunov exponent. Lyapunov exponents look at the behaviour of nearby orbits in a system. They measure the average rates of expansion or contraction by describing the evolution of an infinitesimal ball in phase space. If the principal axes of an ellipsoidal ball are initially $\sigma_i(0)$, then at time $t$ they are $\sigma_i(t) = \sigma_i(0)e^{\lambda_i t}$, where $\lambda_i$ are the Lyapunov exponents

$$\lambda_i = \lim_{t \to \infty} \lim_{\sigma_i(0) \to 0} \frac{1}{t} \ln \frac{\sigma_i(t)}{\sigma_i(0)}. \tag{32}$$

An $n$-dimensional map has $n$ Lyapunov exponents corresponding to $n$ expanding or contracting directions in phase space. The two-dimensional Bogdanov map has two Lyapunov exponents at every point. Since an arbitrary initial direction will tend to expand with the largest principal axis, in almost all cases, choosing a random initial vector $u$ leads to the principal (largest) Lyapunov exponent. The Lyapunov exponents are defined in the limit as the number of iterations of the map goes to infinity, and so they reflect the behaviour on the attractor on which the trajectory ends up, and they filter out the transient behaviour. They are thus the same for points in the same basin of attraction. Chaos is characterized by sensitive dependence on initial conditions; a positive Lyapunov exponent shows exactly this, and is a necessary condition for chaos to occur.

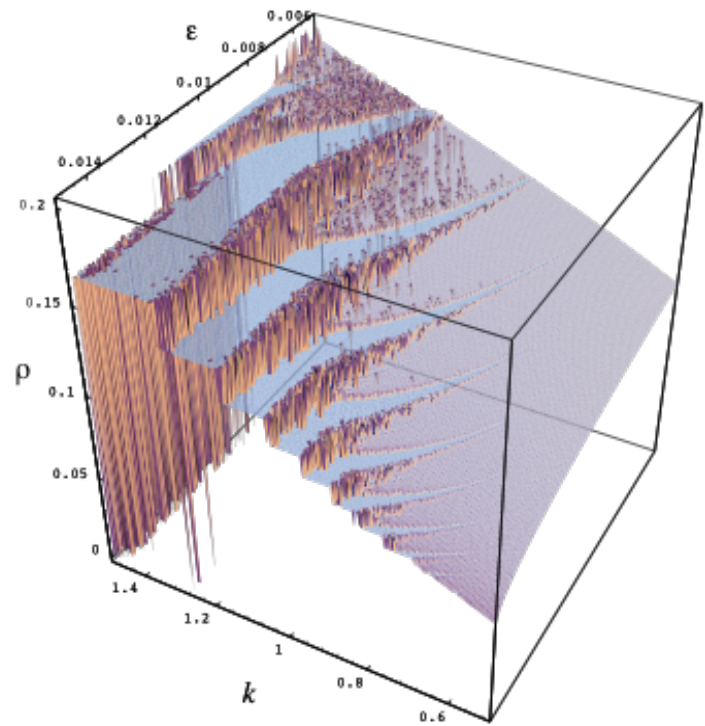

Figure 15: A devil's quarry rotation number plot in the Bogdanov map for $0.005 < \varepsilon < 0.015$ and $0.5 < k < 1.5$ with $\mu = -0.1$. This is an enlargement of part of Fig. 13. Resonances with island chains of orders 2/11, 1/6, 1/7, 1/8, 1/9, 1/10, 1/11, 1/12, and 1/13 are the most prominent. A zero value is plotted at points where iterates escape to infinity.

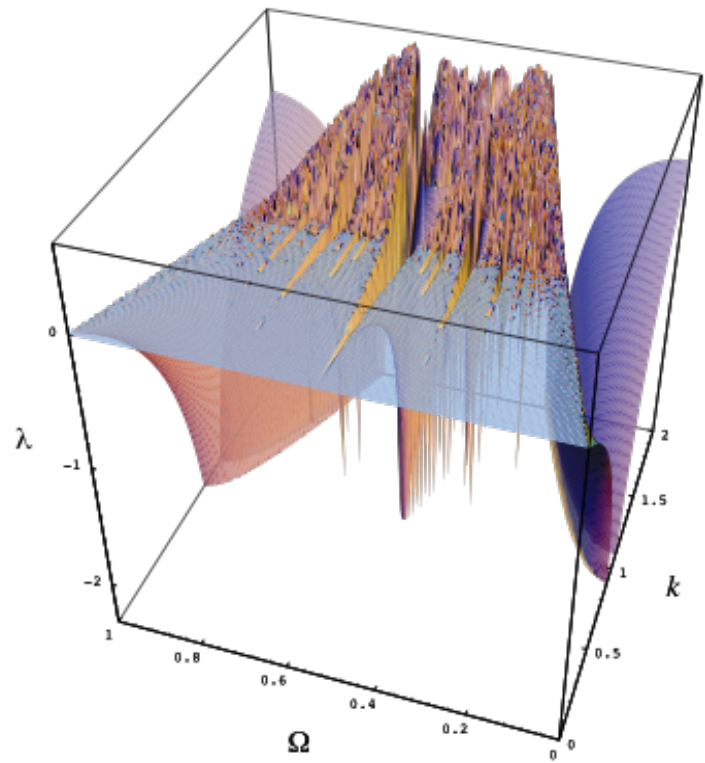

Figure 16: Lyapunov exponent in the sine circle map for $0 < \Omega < 1$ and $0 < k < 2$. Order in the subcritical region $k < 1$ may be contrasted with the chaos of the supercritical region $k > 1$. Notice also the accumulation point of period doubling to chaos that is visible in the larger tongues.

Complementing the devil's quarry of Fig. 14, in Fig. 16 we show Lyapunov exponent in the sine circle map. In this picture we see the sharp contrast between the nonchaotic subcritical region $k < 1$ and the chaos in the supercritical region $k > 1$. Quasiperiodicity outside the tongues shows up as zero Lyapunov exponent. Within the tongues when the map is subcritical, the Lyapunov exponent is negative, and goes to negative infinity at the superstable[10] points in the tongue. As in Fig. 14, we can easily see the accumulation point of period doubling to chaos inside the larger tongues in the circle map. Compare this with Fig. 17, where we show the principal Lyapunov exponent in the Bogdanov map for a fixed initial point with different parameter values in the region $0.005 < \varepsilon < 0.015$ and $0.5 < k < 1.5$ with $\mu = -0.1$, that is, the same parameter values as the devil's quarry rotation number plot of Fig. 15. The resonances are seen as chasms in the picture, as we saw for the circle map in Fig. 16. Outside the resonances, we have quasiperiodic motion which has near zero Lyapunov exponent. The spikes of positive principal Lyapunov exponent are occurring at the edges of resonance, as one would predict from the homoclinic tangles known to exist in these regions, when we have a strange attracting set. The small number of points for which the Lyapunov exponent is positive is a reflection of the fact noted above, that the homoclinic tangles occur over an extremely small interval at the edges of resonance.

## 8. Nonresonant Island Chains

IN a circle map, the one-dimensionality means that there is either a nonresonant invariant circle, or a resonant invariant circle with periodic orbits on it, forming part of an Arnold tongue. Existence of an Arnold tongue in a circle map is synonymous with resonance, as Fig. 14 shows. On the other hand, in a two-dimensional map, an Arnold tongue can coexist with resonant behaviour occurring elsewhere on the phase plane. The continued existence of the island chains outside the interval of resonance with the Hopf circle is shown by comparing Fig. 15 with Fig. 10. The visible plateaux in Fig. 15, which show the extent of resonance with an island chain, are subsets of the Arnold tongues in Fig. 10. We illustrated in Fig. 12 that the island chains in the Bogdanov map continue to exist even when not resonant with the Hopf circle, but are more difficult to detect numerically, having small basins of attraction. The distance between the manifolds of the saddles that contain the foci is small, and most iterates can flow through the nonresonant island chain to end up on the Hopf circle— see the top and bottom illustrations of Fig. 12. We show this in Fig. 18, where we plot rotation number in the Bogdanov map for different initial conditions in the square $(-0.6, -0.7) < (x, y) < (1.0, 0.9)$, at the parameter values $\varepsilon = 0.01$, $\mu = -0.1$, and $k = 1.2$. The picture shows very clearly the existence of several rotation numbers at the same parameter values, and the nonexistence of a rotation number for some initial conditions, which is represented in the picture as rotation number zero. A two-dimensional map like the Bogdanov map need only have a unique rotation number if the attracting set is a circle. This is the case for subcritical circle maps; the rotation number is independent of the initial point, so the map has a unique rotation number for each parameter value. When a circle map is supercritical, it becomes noninvertible and ceases to be a diffeomorphism, and it may possess more than one rotation number, in which case there must be an interval of rotation numbers. In the Bogdanov map, plotting rotation number for different initial points shows that several different rotation numbers can coexist at each parameter value. We can see in Fig. 18 that most initial conditions lead to island chains with rotation numbers $1/7$ or $4/27$, but a few initial conditions at those parameter values lead to the $1/6$ chain which is clearly visible in the picture as six groups of peaks, where the initial conditions are inside the manifolds of the saddles containing the foci for that chain. We can make the comparison of Fig. 18 with Fig. 11 see the positions of the period-6 and period-7 island chains on the phase portrait for these parameter values.

---

[10]At a superstable point in a periodic orbit of period $q$ in a one-dimensional map, $df^q(\theta)/d\theta = 0$, and there is quadratic, rather than linear convergence to the periodic orbit.

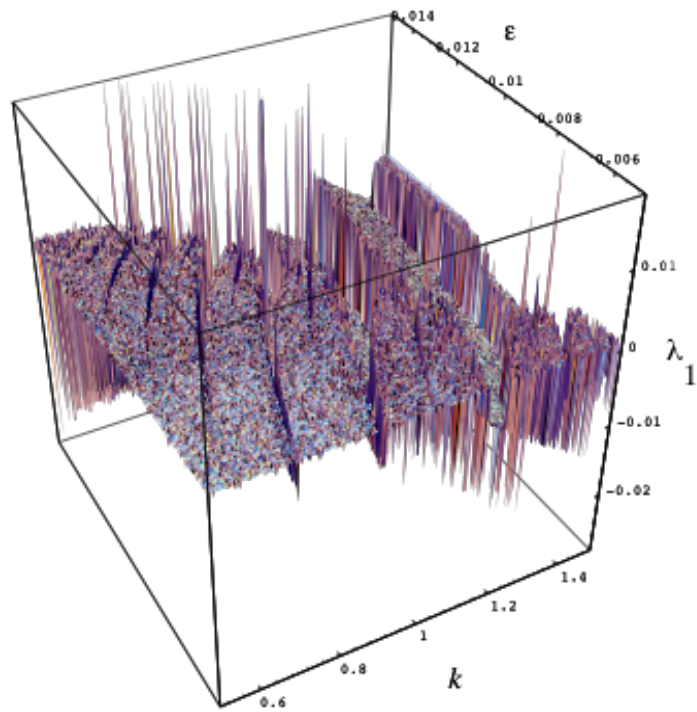

Figure 17: Plot of principal Lyapunov exponent in the Bogdanov map for $0.005 < \varepsilon < 0.015$ and $0.5 < k < 1.5$ with $\mu = -0.1$. Compare this plot with Fig. 15; the resonances that appear as plateaux there are chasms in this picture. A zero value is plotted at points where iterates escape to infinity.

Figure 19 displays the sequence leading up to Fig. 18, and confirms some of the sketches in Fig. 12. In Fig. 19 we show phase portraits of the Bogdanov map for $\mu = -0.1$ with $k = 1.2$, as in Fig. 18, but with $\varepsilon$ varying from $0.002$ to $0.008$. In each portrait we see the part of the attracting set made visible[11] by iterating an initial point close to the origin, plus the period-6 island chain and the associated manifolds of the saddles. In the first portrait, Fig. 19(a), at $\varepsilon = 0.002$, we see the invariant circle well inside the period-6 island chain. The manifolds of the saddles are open, allowing initial points outside the chain to flow through to end up on the Hopf circle. This corresponds to the top sketch in Fig. 12. The next portrait, Fig. 19(b), at $\varepsilon = 0.005$, is at the approximate position of the second sketch in Fig. 12. The invariant circle has moved out and the interval of rotary homoclinic tangles on the inside of the island chain has just ended (remember that this interval is only $\sim 10^{-12}$ in $\varepsilon$ here), so the invariant circle is now resonant with the island chain. In the next portrait, Fig. 19(c), we find the invariant circle in the middle of its interval of resonance at $\varepsilon = 0.0055$. This interval is about to end at $\varepsilon = 0.006$ in Fig. 19(d), as the manifolds of the saddles approach each other on the outside of the island chain, soon to form a rotary homoclinic tangle for a second short interval. The last portrait, Fig. 19(e), shows the situation at $\varepsilon = 0.008$, when the invariant circle has ceased to be in resonance with period 6 and is now on the way out towards period 7. Period 6 still exists but for most initial conditions inside the period-6 island chain, iterates can flow through unimpeded to arrive outside on the invariant circle.

---

[11]Aronson *et al.* [1983] use the term *visible* attractor in this context, and also define its complement, the *invisible* attractor.

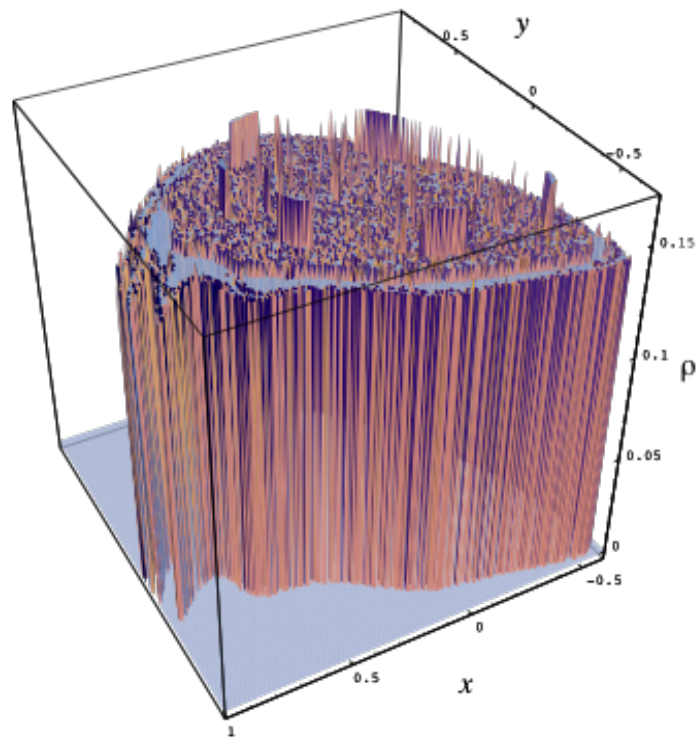

Figure 18: Rotation number in the Bogdanov map for different initial conditions in the square $(-0.6, -0.7) < (x, y) < (1.0, 0.9)$, at the parameter values $\varepsilon = 0.01$, $\mu = -0.1$, and $k = 1.2$. Island chains with rotation numbers $1/7$, $4/27$, and $1/6$ can be seen to coexist. A zero value is plotted at points where iterates escape to infinity.

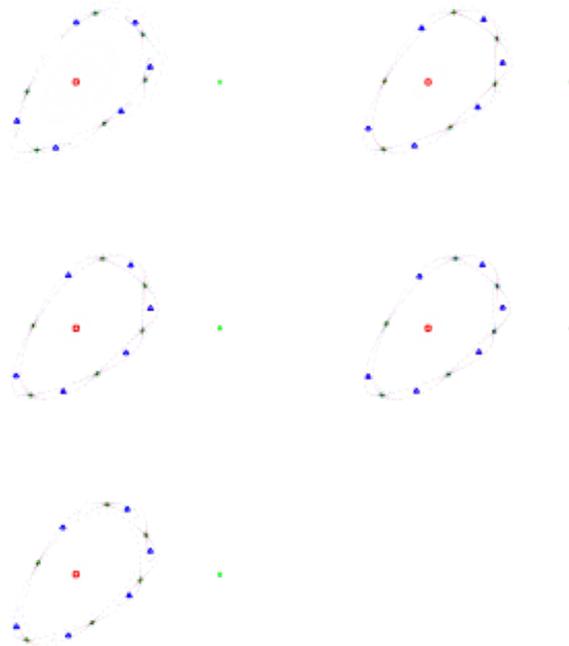

Figure 19: A sequence of phase portraits of the Bogdanov map, varying $\varepsilon$ from 0.002 to 0.008 at $\mu = -0.1$, and $k = 1.2$. This sequence shows the invariant circle moving through resonance with the 1/6 island chain, prior to the situation shown in Fig. 11. The stable (blue) and unstable (red) manifolds of the period-6 saddles show some of the features sketched in Fig. 12. Sinks are marked by blue triangles, sources by red squares, and saddles by green crosses. Iterates of the map for a single initial condition close to the origin are shown in red. The sequence shows, from (a) to (e), scanning from top left to bottom right, phase portraits at $\varepsilon = 0.002$, 0.005, 0.0055, 0.006, and 0.008 respectively. The abscissae are $-0.7 < x < 1.3$, and the ordinates are $-1 < y < 1$.

# 9. The Structure Inside an Arnold Tongue

IT is necessary to understand how the behaviour of the Bogdanov map that we have reported here is reconciled with the demand of the Hopf bifurcation theorem that the bifurcating invariant circle be smooth, and to know whether the same mechanism for the loss of smoothness of an invariant circle, as was first reported by Aronson *et al.* [1983], is also operating in the Bogdanov map. We have noted that the periodic orbits we find in the dissipative Bogdanov map are present continuously as we add dissipation to the area-preserving Bogdanov map. Fixed points and periodic points which were centres in the area-preserving case are perturbed to become foci in the dissipative map. This must be held in mind together with the fact that the Hopf bifurcation theorem requires a smooth invariant circle at the bifurcation, which implies that it must contain nodes when in resonance. The resolution of the apparent paradox lies in looking at the distribution of the Birkhoff periodic orbits in the area-preserving case.

We have seen that periods four and higher are born at the origin in the area-preserving Bogdanov map at $k = 2 - 2\cos(2\pi p/q)$, and exist for $k > 2 - 2\cos(2\pi p/q)$. These periodic orbits correspond to the weak resonances in the dissipative map. We know from the argument using the implicit function theorem which we presented earlier that the periodic points must persist for a ball of $(\varepsilon, \mu)$ values around $(\varepsilon, \mu) = (0, 0)$, and further we know that the eigenvalues must change smoothly so that centres initially become foci. Thus we know that one edge of the $p/q$ tongue must cross the line $\varepsilon = 0$ for $k > 2 - 2\cos(2\pi p/q)$, when it must contain foci. We see from Fig. 10 that the tongues do in fact cross this line. Consider for example the $1/6$ tongue in that figure. It bifurcates from $k = 2 - 2\cos(2\pi/3) = 1$ into the region $\varepsilon > 0$, but one edge bends back so that the tongue crosses the line $\varepsilon = 0$ at a higher value of $k$, say $k_i$. This $k_i$ must be dependent on $\mu$, since we know that at $\mu = 0$, the $1/6$ tongue must exist continuously for $k \geqslant 1$. So when $\mu = 0$, $k_i = 1$ for $p/q = 1/6$, and in general for any tongue, $k_i = 2 - 2\cos(2\pi p/q)$ at $\mu = 0$. As $\mu$ becomes nonzero, the Hopf bifurcation forces $k_i$ to be greater than $2 - 2\cos(2\pi p/q)$, because we know from the Hopf theorem that the $p/q$ tongue is born into the region $\varepsilon > 0$ from a cusp at $k = 2 - 2\cos(2\pi p/q)$, so that the crossing of the line $\varepsilon = 0$ must occur for some $k_i > 2 - 2\cos(2\pi p/q)$. Local to the cusp, the Hopf theorem tells us that the island chains contain nodes, so the foci must begin some distance into the tongue. The Hopf bifurcation also forces the foci to give way to nodes before the edges of the tongue, where they must coalesce with the saddles at saddle–node bifurcations. Thus at any nonzero $\mu$ there must be an interval of nodes at the beginning and around the edges of the tongue. This interval is extremely small as $\mu$ approaches zero.

We can collate all this information about the internal structure of an Arnold tongue in the Boganov map, together with the discussion on this topic by Aronson *et al.* [1983] for the delayed logistic map, to come up with a coherent picture of the behaviour inside a tongue. In Fig. 20 we show our best guess at this point for some of the internal structure of an Arnold tongue in the Bogdanov map. This figure is very distorted and nonlinearly scaled, to enable some of the regions to be distinguished which are in reality minute. The solid lines in Fig. 20 show features which we are certain of; we are less sure about the relative positions of the dotted lines. The two lines AB and A′B are the curves along which the saddle–node bifurcations at the tongue edges occur, meeting in a cusp at B, the beginning of the tongue. Inside the tongue on the curve CC′ the two eigenvalues of the sinks or sources are the same. Remember that we have secondary Hopf bifurcations occurring inside the tongues, which is why we say sinks *or* sources throughout this discussion. On crossing from outside to inside CC′, nodes become foci. On DE, D′E′, GI, and G′I′, rotary homoclinic tangencies occur, and on EF, E′F′, IL, and I′L′, we have rotary heteroclinic tangencies. These rotary homoclinic and rotary heteroclinic tangencies occur on the edges of intervals of rotary homoclinic and rotary heteroclinic tangles respectively. Rotary homoclinic tangles we have already met; in the case of rotary heteroclinic tangles, we have the unstable manifold of a point in a periodic saddle intersecting with the strong stable manifold of the adjacent periodic node in the island chain. Notice that rotary heteroclinic tangles can only occur when the sinks or sources are nodes. JK and J′K′ mark the position of the boundary of differentiability of the invariant circle when the sinks or sources are still nodes; further up the tongue from here, the circle is not even $C^1$.

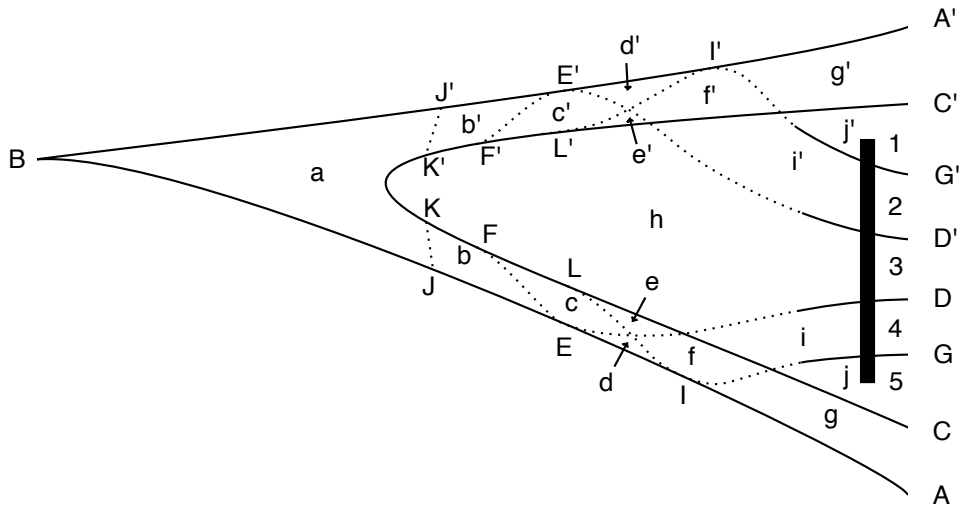

Figure 20: The structure inside an Arnold tongue in the Bogdanov map. The regions (a) to (h) are explained in the text. The vertical line with numbers 1 to 5 next to it represents where the behaviour displayed in the five diagrams of Fig. 12 is to be found in this picture.

In region (a) of Fig. 20, the sinks or sources are nodes, the invariant circle is at least $C^1$, and near the tongue's tip it must be smooth ($C^\infty$). In region (h), on the other hand, the sinks or sources are foci, so the invariant circle is nondifferentiable. In regions (b) and (b′) the sinks or sources are still nodes, but the circle has lost differentiability—it has developed kinks. There are further changes in regions (c) and (c′), where the unstable manifolds of the saddles form rotary heteroclinic tangles with the strong stable manifolds of the nodes. In regions (d) and (d′), this behaviour gives way, after a rotary heteroclinic tangency, to an invariant circle with cusps at the nodes, whereas entering regions (e) and (e′), we have a rotary homoclinic tangency leading to simultaneous rotary homoclinic and heteroclinic tangles. Moving to (f) and (f′), another rotary heteroclinic tangency leads to regions where only the rotary homoclinic tangles remain. Going from (f) to (i), or from (f′) to (i′), the rotary homoclinic tangles are still present, but the nodes turn into foci. Finally, (g) and (g′), and (j) and (j′), are only differentiated because in the former regions, the sinks or sources are nodes, and in the latter regions they are foci. In both cases, moving into these regions from further inside the tongue takes one across a line on which occurs another rotary homoclinic tangency. The invariant circle is completely disconnected from the island chain in regions (g), (g′), (j), and (j′). The island chain still remains in existence up to the saddle–node bifurcation at the edge of the tongue. The boundary of the connection of the island chain to the invariant circle occurs somewhere inside regions (f) and (f′); in other words, for the outer part of these regions, the invariant circle can be disconnected from the homoclinic tangle.

Regions (a) to (h) were first discussed by Aronson *et al.* [1983] for the delayed logistic map. There are some differences between their observations of these regions and ours. They find that in the delayed logistic map, region (g) terminates further along the tongue from its birth with the development of a homoclinic orbit; we have not seen this in the Bogdanov map. We have not observed (a) to (f) for weak dissipation in the Bogdanov map, but the existence of (a) is required by the Hopf theorem, and (b) to (f) to link together (a) and the other regions we have observed; (g), (h), (i), and (j). We have met (h), (i), (i′), (j), and (j′) before; the vertical black line on Fig. 20 represents the position on that diagram of the behaviour drawn in Fig. 12, and the numbers 1 to 5 alongside indicate the numbers of the diagrams from top to bottom in Fig. 12. Regions (h), (j), and (j′) were observed in the sequence of phase portraits of Fig. 19. Figure 19(a) is in (j′), Figs. 19(b)–(d) are in (h), and Fig. 19(e) is in (j). The reason for the difficulty of observing regions (a) to (f) with weak dissipation is that in these regions the sinks or sources are nodes. The proximity of the map to the area-preserving case means that most sinks or sources will be foci. The interval for which nodes are found around the edges of the tongue shrinks to zero as the parameters $\varepsilon$ and $\mu$ approach zero, and these intervals of nodes are already extremely difficult to detect with a computer when $\mu$ is as small as $-0.1$.

There is obviously further structure inside a tongue that is not shown in Fig. 20. In particular, we have concentrated in Fig. 20 on the behaviour of the manifolds of the periodic points in a tongue. Another feature that we must consider is the period-doubling bifurcation. We know that inside the curve of equal eigenvalues in the tongue, within which the sinks or sources are foci, there must be a further curve on which the eigenvalues are equal, but this time negative instead of positive. Inside this curve the eigenvalues are real again, and sinks or sources will become nodes once more. Within this curve is another where one eigenvalue becomes $-1$, leading to a period-doubling bifurcation. The period-$q$ node becomes a period-$q$ saddle, and spawns two period-$2q$ nodes, which undergo the same sequence of bifurcations as we have just described, to become four period-$4q$ nodes, and so on ad infinitum. We saw this period-doubling process in the sine circle map in Figs. 14 and 16. We display the first few steps of the period-doubling process occurring in the $1/6$ tongue of the Bogdanov map in Fig. 21. In this figure, the $1/6$ periodic orbits are plotted on the phase plane for $\varepsilon = 0.01$, $\mu = -0.08$, and $k$ increasing from $1.2$ to $1.6$ in steps of $0.02$. As $k$ increases the $1/6$ island chain moves out, enabling one to separate the behaviour for each $k$ on Fig. 21. The saddle–node bifurcation bringing the island chain into being has occurred at $k$ just less than $1.2$. After an extremely small interval, the nodes, which are unstable, become repelling foci, which then undergo secondary Hopf bifurcations at around $k = 1.21$ to become attracting foci. At about $k = 1.47$, there is another very small interval of nodes, followed by a period-doubling bifurcation, which gives rise to a period-12 orbit, also shown on Fig. 21. From here up to $k = 1.6$, the original period-6 island chain consists of two sets of saddles, and the period-12 island chain is also seen to undergo the same stability change at another period-doubling bifurcation to produce a period-24 island chain, which is not shown. Period doubling is not obvious in the devil's quarry pictures of the Bogdanov map, as it is in the devil's quarry of the sine circle map, because in the Bogdanov map it is nonresonant periodic orbits that are period doubling. The devil's quarry in the Bogdanov map concentrates attention on the resonant behaviour occurring elsewhere in the phase plane. Of course, this cannot be the case in the one-dimensional circle map, where the existence of an Arnold tongue is equivalent to resonance with the periodic orbit within it.

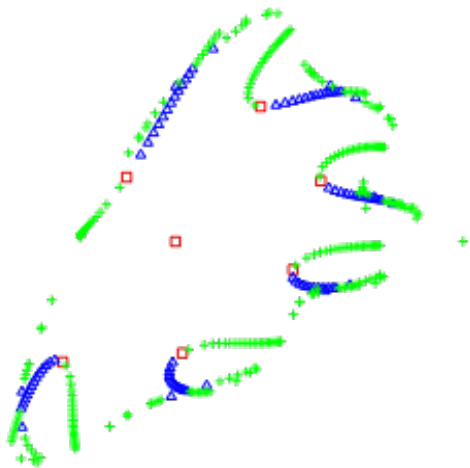

Figure 21: The 1/6 island chain is shown here on the phase plane with $\varepsilon = 0.01$, $\mu = -0.08$, and $k$ increasing from 1.2 to 1.6 in steps of 0.02. As $k$ increases, the outward movement of the island chain enables one to separate the behaviour at different $k$ on the same picture. Moving out, bifurcations of saddle–node, Hopf, and period-doubling type can be picked out immediately. The latter leads to the period-12 orbit also shown here. Sources are displayed as red squares, sinks as blue triangles, and saddles as green crosses. The abscissa is $-0.7 < x < 1.3$, and the ordinate is $-1 < y < 1$.

The other prominent feature inside a tongue missing from Fig. 20, is the secondary Hopf bifurcation. Figure 21 shows a secondary Hopf bifurcation as well as the period-doubling bifurcations that are discussed above. We have already noted the infinite hierarchy of Hopf bifurcations: primary; secondary; tertiary, and so on, that occur in the Bogdanov map. Just as the primary Hopf bifurcation has Arnold tongues associated with it, so too do all the other Hopf bifurcations. These secondary Arnold tongues are linked to the secondary island chains in the area-preserving Bogdanov map in the same way that the primary Arnold tongues are linked to the primary island chains, and the same is true at all levels. Frouzakis *et al.* [1991] have reported secondary Hopf bifurcations and secondary Arnold tongues in a three-dimensional map. They note that the secondary Hopf bifurcations occur on a curve across the primary tongue, meeting the boundary at both sides at Bogdanov–Takens points, and they have observed secondary Arnold tongues emanating from the secondary Hopf bifurcation curve. We have seen the same behaviour here in the Bogdanov map.

We have commented above on the anomalous behaviour of the $1/3$ and 1/4 resonances in the Hénon area-preserving map, and equivalently in the area-preserving Bogdanov map. These become the 1/3 and 1/4 strong resonances in the dissipative case, and Fig. 13 confirms that they are observed in the dissipative Bogdanov map. They retain their peculiar appearances, forming island chains in which the saddles are very much nearer to the origin than are the sinks. We noted above that the 1/3 resonance is the only one in the area-preserving case that can be found for $k < 2 - 2\cos(2\pi/3) = 3$. This remains true when dissipation is added; the 1/3 tongue is the only one that bends round to lower, as well as higher, $k$ values than $2 - 2\cos(2\pi p/q)$.

Note in Fig. 18 the shape of the boundary of the basin of infinity. At these parameter values, the stable and unstable manifolds of the period-1 saddle form a homoclinic tangle very similar to that displayed for the area-preserving Bogdanov map in Fig. 6. The boundary of the basin of infinity follows the closure of the stable manifold of the period-1 saddle. We need to discover how the invariant circle finally disappears as it encounters this boundary.

## 10. Breakup of the Hopf Invariant Circle

JUST as the limit cycle in the Bogdanov vector field is destroyed in a saddle connection, so there is a similar process in the Bogdanov map. For the vector field it is clear by the Poincaré–Bendixson theorem that when the unstable separatrix of the saddle returns within the stable separatrix, there will be a nontrivial $\omega$-limit set, whereas in the opposite case no such set need exist—see Fig. 2. The same is true in the map. However, in the case of a map, when the stable and unstable manifolds, or insets and outsets, of the period-1 saddle do intersect, they do not generically become coincident, but instead cross transversely. The saddle connection is thus replaced by homoclinic tangles in the region of the line $\mu = -7\varepsilon + O(\varepsilon^2)$, and the situation is far more complicated.

We focus now on the bifurcation events which precede the final loss of stability of the invariant circle when we hold the parameters $k$ and $\mu$ fixed, and increase the Hopf parameter, $\varepsilon$. We start by considering $k = 1.2$. Figures 15 and 19 indicate that at $\mu = -0.1$, as we increase $\varepsilon$ from zero, we observe first an interval of period-6 resonance, followed by an interval of period-7 resonance. The bifurcation diagram of Fig. 22 demonstrates that this is the case over a broad range of $\mu$. Between the intervals of resonance with rotation numbers 1/6 and 1/7 we observe phase locking with an orbit of rotation number 2/13. Consistent with the Farey tree ordering of rationals, between 1/6 and 2/13 there is a small interval of resonance with rotation number 3/19; similarly between 2/13 and 1/7 we observe resonance with an orbit of rotation number 3/20, and so on. Also shown in the bifurcation diagram is the approximate location of the bifurcation arc corresponding to the period-1 saddle being homoclinic, as we described above. The bifurcation arc denoting the end of period-7 resonance crosses this arc at an intermediate value of $\mu$ on the diagram. Thus, for small negative $\mu$, once the invariant circle has entered resonance with the period-7 point, it does not reappear as $\varepsilon$ is increased. In contrast, for larger negative $\mu$, under increasing $\varepsilon$, there is a clearly defined interval of period-7 resonance bounded by intervals of nonresonant behaviour, and by other resonances. The bifurcation diagram of Fig. 22 may be compared with Fig. 23, which shows rotation number for the same parameter range as Fig. 22, and displays plateaux at the rotation numbers mentioned above.

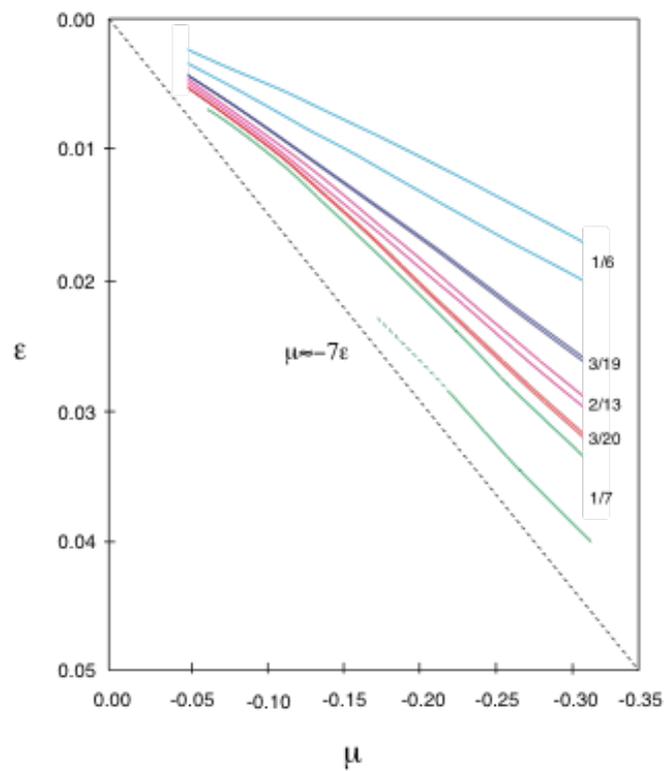

Figure 22: Bifurcation diagram of the Bogdanov map for the parameter range $0 < \varepsilon < 0.05$ and $-0.35 < \mu < 0$ with $k = 1.2$. The boundaries of the 1/7 (green), 3/20 (red), 2/13 (magenta), 3/19 (blue), and 1/6 (cyan) tongues are shown together with the approximate location (dotted) of homoclinicity of the period-1 saddle: $\mu = -7\varepsilon + O(\varepsilon^2)$.

Fixing $\mu = -0.1$, we show a sequence of phase portraits as $\varepsilon$ is increased from zero. These portraits were produced using a cell-to-cell mapping algorithm; see for example Hsu [1987]. The first portrait, Fig. 24(a), at $\varepsilon = 0.0025$ shows the invariant circle, coloured magenta, together with its basin of attraction, coloured yellow. Also in this portrait are two periodic attractors, image points of which are marked with black dots. The basin of the period-6 attractor is coloured cyan; that of the period-7 attractor, red. Starting conditions which diverge to infinity are coloured blue. The boundary of the basin of infinity is defined by the closure of the stable manifold of the period-1 saddle. This basin boundary in general may either be smooth or fractal. The basin boundary undergoes a *smooth–fractal* basin bifurcation when the stable and unstable manifolds of the period-1 saddle develop a homoclinic tangency. The basin of the period-6 attractor is defined by the closure of the stable manifold of the period-6 saddle; similarly, the basin of the period-7 attractor is defined by the closure of the stable manifold of the period-7 saddle. In Fig. 24(a), the basin boundary of the invariant circle is thus delineated by the union of the closures of the insets of the period-1, period-6, and period-7 saddles.

As $\varepsilon$ is increased the invariant circle grows and, as Fig. 24(b) shows, at $\varepsilon = 0.0045$ it approaches the basin of the period-6 attractor. By $\varepsilon = 0.005$, the invariant circle is in resonance with the period-6 solution, Fig. 24(c). The rotary homoclinic intersection of the inner branches of the manifolds of the period-6 saddle provides the topological signature of entrainment here, as we noted in discussing Fig. 19. Additionally, at the beginning of the interval of resonance, the distance between the image points of the period-6 saddle and the attractor tends to zero. Between $\varepsilon = 0.005$ and $\varepsilon = 0.0075$, the interval of period-6 resonance ends. Associated with this, the outer branches of the manifolds of the period-6 saddle undergo a rotary homoclinic tangency and the distance between the attractor and the period-6 saddle increases from zero. Whilst in the interval of period-6 resonance we may view the closure of the unstable manifolds of the period-6 saddle as an attracting set; the attractor however is the period-6 solution. The portrait at $\varepsilon = 0.0075$, Fig. 24(d), shows once more period-6 and period-7 solutions coexisting with the invariant circle. Now however, due to the second rotary homoclinic tangency of the period-6 invariant manifolds, period 6 is located in the interior of the invariant circle. As before, the basin boundary of the invariant circle is defined by the union of the closures of the stable manifolds of the period-1, period-6 and period-7 saddles. We note also that between $\varepsilon = 0.005$ and $\varepsilon = 0.0075$ the period-1 saddle develops a homoclinic tangency; the portrait at $\varepsilon = 0.0075$ clearly shows thin fractal layers of the blue basin of infinity within the yellow basin of the invariant circle. The basin boundary of the finite attractor has therefore undergone a smooth–fractal basin bifurcation.

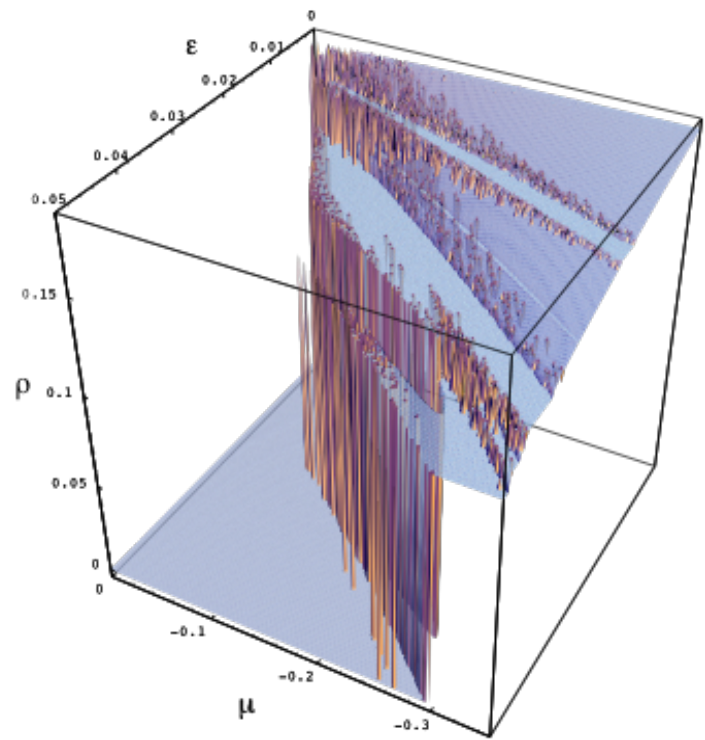

Figure 23: Rotation number in the Bogdanov map for $0 < \varepsilon < 0.05$ and $-0.35 < \mu < 0$ with $k = 1.2$. A zero value is plotted at points where iterates escape to infinity. Resonance with rotation numbers $1/6$, $1/7$, $2/15$, and $1/8$ is most obvious. This figure should be compared with the bifurcation diagram of Fig. 22.

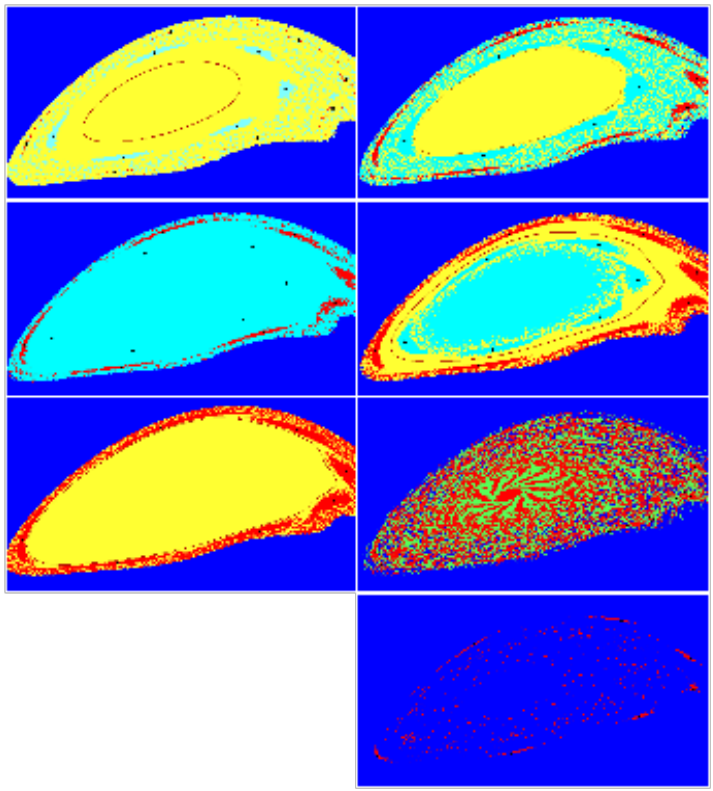

Figure 24: A sequence of phase portraits of the Bogdanov map, varying $\varepsilon$ at $\mu = -0.1$, and $k = 1.2$. The invariant circle is shown in magenta, and its basin of attraction is yellow. Black dots mark periodic attractors. The basin of period 6 is coloured cyan and that of period 7 is red. Green represents the basin of period 42. Initial points which diverge to infinity are in blue. The sequence shows, from (a) to (h), scanning from top left to bottom right, phase portraits at $\varepsilon = 0.0025$, 0.0045, 0.005, 0.0075, 0.01, 0.0125, 0.015, and 0.0194 respectively. The abscissae are $-0.6 < x < 0.8$, and the ordinates are $-0.8 < y < 0.8$.

As we expect from the bifurcation diagram of Fig. 22, increasing $\varepsilon$ further results in the invariant circle approaching the period-7 basin boundary, as the portrait at $\varepsilon = 0.01$, Fig. 24(e), shows. The period-6 attractor was not detected numerically here, probably due to the finite cell size used, since Fig. 19 shows that period 6 does exist at this point. A further small increase in $\varepsilon$ results in the distance between the period-7 saddles and the attractor tending to zero, and there is an associated period-7 rotary homoclinic tangency as the attractor enters period-7 resonance. However, in contrast with the period-6 phase locking, we do not observe that the attractor reappears at any subsequent value of $\varepsilon$. As the portraits at $\varepsilon = 0.015$ and $\varepsilon = 0.0194$ show, Figs. 24(g) and 24(h) respectively, the final attractor we observe at these values of $k$ and $\mu$ is the period-7 point. At these parameter values therefore the invariant circle finally loses stability when simultaneously the period-7 saddle develops a rotary homoclinic tangency and the distance between the attractor and the period-7 saddle tends to zero.

In order to understand why this should be the case, it is helpful to first consider the origin of the periodic orbits we observe in the context of the global bifurcation associated with the homoclinic tangency of the period-1 saddle. Once this tangency progresses to a homoclinic intersection, the Smale–Birkhoff homoclinic theorem proves that horseshoe dynamics govern the local bifurcations that take place; infinitely many periodic orbits will be present. However here we have observed periodic orbits *prior* to the first homoclinic tangency of the period-1 saddle. The creation of horseshoes and periodic orbits before a homoclinic tangency was explained by Gavrilov & Shilnikov [1972; 1973]. By constructing a return map in a small rectangle near the location of a homoclinic tangency in the phase plane, they show that the critical value, $\varepsilon_c$, of the control parameter $\varepsilon$, at which the tangency occurs may be *inaccessible* as it is approached from either $\varepsilon > \varepsilon_c$, or $\varepsilon < \varepsilon_c$, or both. By inaccessible, they mean that as $\varepsilon$ approaches $\varepsilon_c$, an infinite sequence of saddle–node bifurcations, or folds, are encountered. Thus the global bifurcation may be viewed as the accumulation of an infinite sequence of local bifurcations. The type of homoclinic tangency we observe for the period-1 saddle here permits a sequence of subharmonic folds to take place for $\varepsilon < \varepsilon_c$, and so we may associate the period-6 and period-7 orbits with an infinite increasing sequence of subharmonics which accumulate on the first homoclinic tangency and which are created before the period-1 saddle is homoclinic. This infinite increasing sequence of subharmonics accumulates on the first homoclinic tangency of the period-1 saddle in two senses: the location of image points of subharmonics in the phase plane is such that orbits of successively higher period are seen to approach the tangencies between the stable and unstable manifolds of the period-1 saddle. Also, as the parameter $\varepsilon$ is increased towards the value at which the period-1 saddle is homoclinic, orbits of successively higher period are created in local saddle–node bifurcations.

We can consider the periodic orbits we observe in this context in the light of a theorem of Alligood *et al.* [1991]. The theorem explains the smooth–fractal basin bifurcation associated with the first homoclinic tangency of the period-1 saddle in the dissipative Hénon map. Alligood *et al.* have specifically investigated the role of the increasing sequence of saddles in this case and prove that once a homoclinic intersection develops, the basin boundary must necessarily jump inwards and accumulate on the stable manifold of some subharmonic saddle, where this subharmonic is created prior to the first tangency in a local fold bifurcation. Once this smooth–fractal basin bifurcation has taken place, in principle a finite sequence of fractal–fractal basin bifurcations may occur in which the stable manifold of the period-1 saddle solution jumps inwards in a series of steps, accumulating on the stable manifolds of interior members of the increasing sequence of saddles (*interior* in the sense that they are further from the basin boundary in the phase plane). Such fractal–fractal basin bifurcations have been observed and explained for a wide variety of planar dynamical systems, for example, the dissipative Hénon map by Grebogi *et al.* [1987], Alligood & Sauer [1988], and Eschenazi *et al.* [1989], and the Duffing oscillator by Lansbury *et al.* [1992].

We may view the final loss of stability of the invariant circle in the Bogdanov map as being caused by two competing sequences of bifurcations: as we increase $\varepsilon$, the invariant circle passes through resonance with a series of periodic orbits whilst growing out to become the closure of the unstable manifold of the period-1 saddle. While the attractor grows, the basin boundary first undergoes a smooth–fractal basin bifurcation, then jumps inwards in a series of steps, approaching the attractor. The phase portrait at $\varepsilon = 0.015$, Fig. 24(g), shows the stable manifold of the period-1 saddle accumulating on the stable manifold of the period-7 saddle. The closure of the stable manifold of the period-7 saddle is therefore numerically equivalent to the closure of the stable manifold of the period-1 saddle. Starting conditions on one side of this manifold are contained in the basin of the period-7 solution; all starting conditions on the other side escape to infinity, and it is not therefore possible for the invariant circle to reappear.

At larger negative values of $\mu$, the rotary homoclinic tangency of the outward directed branches of the manifolds of the period-7 saddle, signalling the reappearance of the invariant circle, takes place before the basin boundary jumps inward to accumulate on the inset of the period-7 saddle. This is why at larger values of $\mu$, we observe an interval of resonance with rotation number $1/7$. The phase portrait at $\varepsilon = 0.0125$, Fig. 24(f), is interesting in that it shows a period-doubled period-3 solution, the green basin in the red basin of each period-7 image, which has rotation number $6/42$, and does not form part of the Farey sequence associated with the period-1 saddle. The absence of the invariant circle in this portrait indicates that the first period-7 rotary homoclinic tangency has taken place. Comparing this portrait with that at $\varepsilon = 0.015$, Fig. 24(g), we see that by this value of $\varepsilon$, the period-42 attractor has lost stability and the area formerly in its basin is now in the basin of infinity.

The parameter $k$ determines the rotation number of the resonances which dominate in the Bogdanov map. If we increase $k$ to 1.44, we observe phase locking with rotation numbers $1/5$ and $1/6$, as Fig. 25 shows. Fixing $\mu = -0.1$, a similar sequence of bifurcations to the one we have described above is observed while increasing $\varepsilon$ from zero. In Fig. 26(a) we show a phase portrait at $\varepsilon = 0.0001$. The basin of the invariant circle is shown coloured yellow. In addition there is a period-5 attractor, the basin of which is coloured red, and a period-6 attractor, the basin of which is coloured green. The basin of infinity, coloured blue, is observed to layer on the green basin of the period-6 attractor. Thus we surmise that the closure of the stable manifold of the period-6 saddle is numerically equivalent to the closure of the stable manifold of the period-1 saddle. Increasing $\varepsilon$, the invariant circle grows in size and approaches the period-5 saddle, Fig. 26(b). In this portrait, a periodic attractor of rotation number $3/18$ makes its appearance, together with its basin, coloured magenta. This period-18 attractor is part of the sequence of subharmonics accumulating on the homoclinic tangency of the period-6 saddle. The portrait at $\varepsilon = 0.0011$, Fig. 26(c), shows the attractor just before the interval of period-5 resonance takes place. The next phase portrait we show is at $\varepsilon = 0.0032$, Fig. 26(d). The invariant circle has by now passed through resonance with the period-5 solution. The period-5 solution has undergone a secondary supercritical Hopf bifurcation, as shown by the five small invariant circles, the basins of which are coloured pink. A further small increase in $\varepsilon$ results in these invariant circles touching their basin boundaries, the yellow–pink boundary, and disappearing in a blue-sky catastrophe. Increasing $\varepsilon$ further, the invariant circle continues to grow in size; the portrait at $\varepsilon = 0.0049$, Fig. 26(e), shows the attractor approaching the brown basin of a subharmonic of rotation number $2/11$, the appearance of this periodic orbit being consistent with the Farey ordering of the rationals.

## 11. Conclusions

THE Bogdanov map has an extremely rich dynamical structure produced by the interaction of Hamiltonian and dissipative dynamics. The structure is at once complicated and simple; complicated because it is infinite in quantity, and yet simple in that it is self-similar on all scales except the topmost. Apart from the primary level of structure, where the fixed points are constrained to lie a constant distance apart, at all other scales one sees the creation and annihilation of Birkhoff periodic orbits at saddle–node bifurcations, forming Arnold tongues which are born at Hopf bifurcations. The infinite structure of Birkhoff periodic orbits is a legacy of the area-preserving limit of the Bogdanov map. This Hamiltonian limit imposes great constraints on the dissipative dynamics. The Arnold tongues in the dissipative map are forced to bend back on themselves to recross the Hopf bifurcation line on which they were born in order to accommodate the requirements of the Hamiltonian structure.

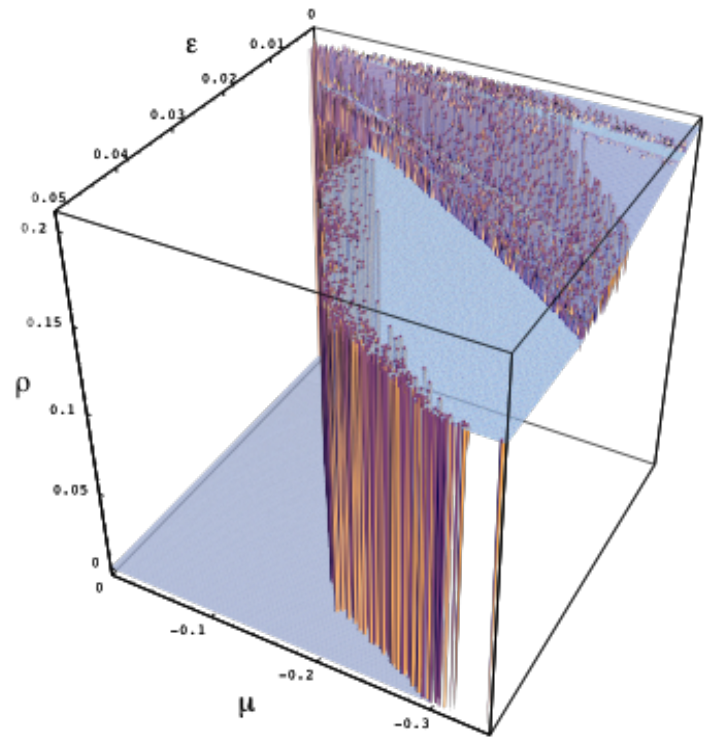

Figure 25: Rotation number in the Bogdanov map for $0 < \varepsilon < 0.05$ and $-0.35 < \mu < 0$ with $k = 1.44$. A zero value is plotted at points where iterates escape to infinity. Notice that phase locking with rotation numbers $1/5$, $2/11$, and $1/6$ is most prominent at this value of $k$.

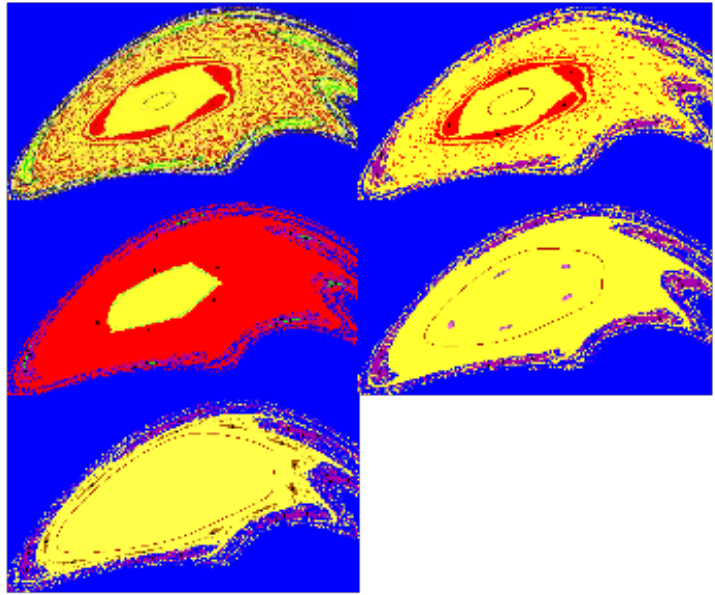

Figure 26: A sequence of phase portraits of the Bogdanov map, varying $\varepsilon$ at $\mu = -0.1$, and $k = 1.44$. The invariant circle is shown in magenta, and its basin of attraction is yellow. Black dots mark periodic attractors. The basin of period 5 is coloured red, that of period 6 is green, and points diverging to infinity are coloured blue. Pink represents the basins of attraction of the invariant circles formed at the secondary Hopf bifurcation of the period-5 sinks. Magenta and brown represent the basins of periods 18 and 11. The sequence shows, from (a) to (e), scanning from top left to bottom right, phase portraits at $\varepsilon = 0.0001$, 0.0002, 0.0011, 0.0032, and 0.0049 respectively. The abscissae are $-0.6 < x < 0.8$, and the ordinates are $-0.8 < y < 0.8$.

The relationship between the tongues and the invariant circles born at the Hopf bifurcations is also complex. The Hopf circles, in passing through the island chains which comprise the tongues, form complicated Birkhoff attracting sets which have an interval of rotation numbers. These strange attracting sets exist for very small parameter ranges on the edges of resonance with an island chain, due to rotary homoclinic tangles of the manifolds of the saddles in the chain. The region of resonance with an island chain is itself a subset of the region for which the island chain exists. The resonant regions of the tongues form what we have termed a *devil's quarry* in the parameter space; a reference to the devil's staircase in circle maps, since both are built with the Farey tree structure. The island chains still exist for some distance beyond this resonant region, but are disconnected from the Hopf circle.

It is the dissipative Hopf parameter $\varepsilon$ which determines the birth and growth from the origin of the primary Hopf invariant circle, the stability of which is set by $\mu$. The Hamiltonian discretization parameter $k$ determines the birth and growth of the island chains. Chaos becomes more evident with increasing $k$. Small $k$ provides a good approximation to the vector field from which the Bogdanov map is derived, and there is increasingly wild separatrix behaviour as $k$ increases. Altering the two parameters $\varepsilon$ and $k$ allows resonance or otherwise of the invariant circle with the infinity of island chains. Changing both parameters simultaneously allows one to plot a course into the tongues leading from quasiperiodicity to periodicity to chaos, or from quasiperiodicity to chaos, depending on the route chosen. It is the complicated internal structure of the tongues that determines which of these scenarios will occur. This internal structure is linked to tongue overlap; when the tongues overlap, they accumulate on the boundaries of the resonances within other tongues in the Farey sequence. Before the rotary homoclinic tangencies at the edges of resonance there is an accumulation of periodic points, and the attractor must resonate with all of these as it grows out to become the closure of the unstable manifolds. Gavrilov & Shilnikov [1972; 1973] have explained the creation of horseshoes and periodic orbits before a homoclinic tangency, and this clarifies the situation in the Bogdanov map. The correspondence between tongue overlap and the loss of smoothness of the invariant circle is of great interest [Aronson *et al.*, 1983; Bohr *et al.*, 1984].

The eventual destruction of the Hopf circles occurs as they grow outwards and their basin boundaries jump inwards in a series of steps, accumulating on the stable manifolds of subharmonic saddles. There is a smooth–fractal basin bifurcation at the first rotary homoclinic tangency, followed by a finite series of fractal–fractal basin bifurcations. Similar bifurcation sequences have been observed in other systems; by Grebogi *et al.* [1987], Alligood & Sauer [1988], and Eschenazi *et al.* [1989] for the dissipative Hénon map, and by Lansbury *et al.* [1992] for the Duffing oscillator.

After a numerical study of the Fermi map in the weakly-dissipative regime, Lieberman & Tsang [1985; 1986] concluded that persistent chaotic motion, as opposed to transient chaos, does not exist continuously when adding dissipation to an area-preserving twist map. On the other hand, Schmidt & Wang [1985] and Chen *et al.* [1986] found that persistent chaos does exist continuously in the Hamiltonian limit of the dissipative standard map, and of a class of maps with constant Jacobian. Casdagli [1988] has also investigated the dissipative standard map. He found that strange attractors are rare with weak dissipation. He provides a theorem to show that no strange attractors with rotational chaos can be found in dissipative twist maps for weak enough dissipation. In the Bogdanov map, plots of the Lyapunov exponent show that persistent chaos is far rarer than transient chaos for weak dissipation. We know however that persistent chaos must exist continuously as the dissipation tends to zero in the Bogdanov map. Consider Fig. 2, which shows the saddle-connection in the Bogdanov vector field on the line $u_2 = -7u_1 + O(u_1^2)$. As we have discussed previously, this saddle-connection line becomes broadened into a wedge of homoclinic tangles in the Bogdanov map. These exist about the line $\mu = -7\varepsilon + O(\varepsilon^2)$, so we know that chaos exists continuously for any $k$ as $\varepsilon$ and $\mu$ tend to zero within this wedge.

Some of the behaviour of the Bogdanov map was first reported in another two-dimensional map by Aronson *et al.* [1983]. The fundamental difference between their model and ours is that their system, the delayed logistic map, is never globally Hamiltonian, so it does not have the infinite hierarchy of Hamiltonian structure that the Bogdanov map inherits. Nevertheless, we find that the structure inside a tongue in the Bogdanov map supports that found for example by Aronson *et al.* [1983] and Frouzakis *et al.* [1991] in other maps, and by Aronson *et al.* [1986], Schreiber *et al.* [1988], and Peckham [1990] in models of Poincaré maps of periodically forced oscillators. Thus we believe that we have achieved what we set out to do; we have produced a good model for the dynamics of periodically forced oscillators. At the same time, the Bogdanov map is also a useful system in which to observe the effect of dissipative perturbations on Hamiltonian structure.

## 12. Acknowledgements

JHEC wishes to thank Nik Burić, who helped greatly in the clarification of various points, Oreste Piro with whom he had many useful conversations, and Emilio Hernández for help with obtaining some of the figures. ANL would like to thank Bruce Stewart for useful discussions. We should like to acknowledge the financial support of the Science and Engineering Research Council (SERC). JHEC is also pleased to thank the AEJMC Foundation for support.

## A.1. Normal Form Calculation—Stability Index

IN In order to distinguish between subcritical and supercritical Hopf bifurcations, we need to perform a standard normal form calculation. We follow Lauwerier [1986], and write the map in the form

$$\begin{aligned} x' &= y, \\ y' &= Ax + By + \sum_{\substack{j,k \\ j+k \geq 2}} g_{jk} x^j y^k, \end{aligned} \tag{33}$$

so that the fixed point is at the origin. The Jacobian is then

$$\begin{pmatrix} 0 & 1 \\ A & B \end{pmatrix} \tag{34}$$

with determinant $\lambda \bar{\lambda} = -A$, and trace $\lambda + \bar{\lambda} = B$ so the eigenvalues $\lambda$ and $\bar{\lambda}$ are

$$\frac{B}{2} \pm \sqrt{\left(\frac{B}{2}\right)^2 + A} = (1+\nu)e^{\pm i\alpha}. \tag{35}$$

This means that the eigenvalues cross the unit circle $|\lambda| = 1$ when $\nu$ passes through zero; then $\cos\alpha = B/2$ and $\sin\alpha = \sqrt{(-A - B^2/4)}$. Now we introduce complex coordinates

$$\begin{aligned} z &= ix\bar{\lambda} - iy, \\ \bar{z} &= -ix\lambda + iy, \end{aligned} \tag{36}$$

so that on the unit circle $2x \sin\alpha = z + \bar{z}$ and $2y \sin\alpha = \lambda z + \bar{\lambda}\bar{z}$. We write the map in the complex form above to obtain

$$\begin{aligned} z' &= ix'\bar{\lambda} - iy' \\ &= \lambda z + \sum_{\substack{j,k \\ j+k \geq 2}} a_{jk} z^j \bar{z}^k. \end{aligned} \tag{37}$$

This is the prenormal form, where the $a_{jk}$ are functions of the $g_{jk}$, $\lambda$, $\bar{\lambda}$, and $\alpha$. On the unit circle

$$a_{20} = -i \frac{g_{20} + \lambda g_{11} + \lambda^2 g_{02}}{4 \sin^2 \alpha}, \tag{38}$$

$$a_{11} = -i \frac{g_{20} + g_{11} \cos\alpha + g_{02}}{2 \sin^2 \alpha}, \tag{39}$$

$$a_{02} = -i \frac{g_{20} + \bar{\lambda} g_{11} + \bar{\lambda}^2 g_{02}}{4 \sin^2 \alpha}, \tag{40}$$

$$a_{21} = -i \frac{3g_{30} + (2\lambda + \bar{\lambda}) g_{21} + (2 + \lambda^2) g_{12} + 3\lambda g_{03}}{8 \sin^3 \alpha}. \tag{41}$$

Following Arnold [1983], we define new complex coordinates $w$ and $\bar{w}$, where

$$w = z + \sum_{\substack{j,k \\ j+k \geq 2}} p_{jk} z^j \bar{z}^k, \tag{42}$$

such that most nonlinear terms can be removed. The map will then be in the normal form

$$w' = \lambda w + \sum_{\substack{j,k \\ j+k \geq 2}} n_{jk} w^j \bar{w}^k. \tag{43}$$

We can substitute for $w$ from Eq.(42) and compare the result with the iterated form of the same equation, which is

$$w' = z' + \sum_{\substack{j,k \\ j+k \geq 2}} p_{jk} z'^j \bar{z}'^k. \tag{44}$$

We substitute for $z'$ above using Eq.(37) and compare coefficients of $z^j \bar{z}^k$ on the unit circle to obtain these expressions:

$$n_{20} - p_{20}\lambda(\lambda - 1) = a_{20}, \tag{45}$$

$$n_{11} + p_{11}(\lambda - 1) = a_{11}, \tag{46}$$

$$n_{02} + p_{02}\bar{\lambda}^2(\lambda^3 - 1) = a_{02}, \tag{47}$$

$$n_{30} - p_{30}\lambda(\lambda^2 - 1) = -\bar{p}_{02}n_{11} - 2n_{20}p_{20} + \lambda(2a_{20}p_{20} + \bar{a}_{02}p_{11})$$
$$+ a_{30}, \tag{48}$$

$$n_{21} = -(\bar{p}_{11} + p_{20})n_{11} - 2\bar{p}_{02}n_{02} - 2p_{11}n_{20}$$
$$+ \bar{\lambda}(p_{11}a_{20} + 2p_{20}\bar{a}_{02}) + \lambda(2p_{20}a_{11} + p_{11}\bar{a}_{11})$$
$$+ a_{21}, \tag{49}$$

$$n_{12} + p_{12}\bar{\lambda}(\lambda^2 - 1) = -(\bar{p}_{20} + p_{11})n_{11} - 2\bar{p}_{11}n_{02} - 2p_{02}n_{20}$$
$$+ \bar{\lambda}(p_{11}a_{11} + 2p_{02}\bar{a}_{11}) + \lambda(2p_{20}a_{02} + p_{11}\bar{a}_{20})$$
$$+ a_{12}, \tag{50}$$

$$n_{03} + p_{03}\bar{\lambda}^3(\lambda^4 - 1) = -2\bar{p}_{20}n_{02} - 2p_{02}n_{11} + \bar{\lambda}(2p_{02}\bar{a}_{20} + p_{11}a_{02})$$
$$+ a_{03}. \tag{51}$$

We aim to have as many $n_{ij}$ as possible set equal to zero. Since $\lambda \neq 1$ and $\lambda^3 \neq 1$ we can remove all the quadratic terms by setting

$$p_{20} = \frac{a_{20}}{\lambda - \lambda^2}, \tag{52}$$

$$p_{11} = \frac{a_{11}}{\lambda - 1}, \tag{53}$$

$$p_{02} = \frac{a_{02}}{\lambda - \bar{\lambda}^2}. \tag{54}$$

We can remove $n_{30}$, $n_{12}$, and $n_{30}$ in the same fashion, since $\lambda^2 \neq 1$ and $\lambda^4 \neq 1$. However, the last cubic term, $n_{21}$, cannot be removed and we have

$$n_{21} = \frac{|a_{11}|^2}{1 - \bar{\lambda}} + \frac{2|a_{02}|^2}{\lambda^2 - \bar{\lambda}} + \frac{2\lambda - 1}{\lambda(1 - \lambda)} a_{11} a_{20} + a_{21}. \tag{55}$$

In general, the coefficient of the $p_{jk}$ term in the expansion is $\lambda - \lambda^j \bar{\lambda}^k$, so on the unit circle, if $\alpha/2\pi$ is irrational so that $\lambda^q \neq 1$ for any integer $q$, the only $n_{jk}$ terms that we cannot set equal to zero are $n_{21}$, $n_{32}$, $n_{43}$, etc., corresponding to $w^2\bar{w}$, $w^3\bar{w}^2$, $w^4\bar{w}^3$, etc. in the normal form. Thus

$$w' = \lambda w + n_{21}w^2\bar{w} + O(|w|^5) \tag{56}$$

is the normal form when $\alpha/2\pi$ is irrational. If $\alpha/2\pi$ is rational and of the form $p/q$, where $p/q$ is in its lowest terms and $q > 4$, we have a weak resonance and additional terms appear in the normal form. From the coefficient of the $p_{jk}$ term we can see that $n_{0q-1}$ is the first additional term, because $\lambda - \bar{\lambda}^{q-1} = 0$ if $\lambda^q = 1$. Thus Eq.(56) is the normal form for a weak resonance also (but $O(|w|^4)$ replaces $O(|w|^5)$ in Eq.(56) if $q = 5$). The cases $q = 1, 2, 3$ or $4$ are the so-called strong resonances specifically excluded from the Hopf theorem. The $q = 1$ and $q = 2$ cases correspond to double $1$ eigenvalues (the Bogdanov–Takens point for maps) and double $-1$ eigenvalues respectively. The normal form for $q = 3$ is

$$w' = \lambda w + n_{02}\bar{w}^2 + n_{21}w^2\bar{w} + O(|w|^4), \tag{57}$$

and for $q = 4$ we have the normal form

$$w' = \lambda w + n_{21}w^2\bar{w} + n_{03}\bar{w}^3 + O(|w|^5). \tag{58}$$

There are still some unsolved problems associated with the behaviour of the strong resonances [Arnold, 1983; Whitley, 1983; Arrowsmith & Place, 1990b].

We now introduce polar coordinates $w = re^{i\theta}$ and substitute $n_{21} = be^{i\gamma}$ and $\lambda = (1+\nu)e^{i\alpha}$ into Eq.(56) to obtain

$$r' = r|(1+\nu)e^{i\alpha} + be^{i\gamma}r^2| + \cdots. \tag{59}$$

In the lowest order approximation, this gives

$$r'^2 = (1+2\nu)r^2 + 2b\cos(\alpha-\gamma)r^4. \tag{60}$$

This is a one-dimensional map with fixed points $r = 0$ and

$$r = \sqrt{-\frac{\nu}{b\cos(\alpha-\gamma)}}. \tag{61}$$

The latter exists if $\nu < 0$ and $b\cos(\alpha - \gamma) > 0$, when it is unstable, or if $\nu > 0$ and $b\cos(\alpha-\gamma) < 0$, when it is stable. Thus we get one of two possible pictures (see Fig. 7): the former case is called a subcritical Hopf bifurcation, and the latter a supercritical Hopf bifurcation. $b\cos(\alpha - \gamma) = \text{Re}(\bar{\lambda}n_{21})$ is the stability index that distinguishes between the two cases, and is given by

$$\text{Re}(\bar{\lambda}n_{21}) = \text{Re}(\bar{\lambda}a_{21}) - |a_{02}|^2 - \frac{1}{2}|a_{11}|^2 - \text{Re}\left(\frac{(1-2\lambda)\bar{\lambda}^2}{1-\lambda}a_{11}a_{20}\right). \tag{62}$$

(This is also obtained in [Iooss, 1979; Whitley, 1983; Lauwerier, 1986; Arrowsmith & Place, 1990b].)

Putting the Bogdanov map in the form of Eq.(33), we have

$$\begin{array}{rcl} x' & = & \hat{y}, \\ \hat{y}' & = & -(1+\varepsilon)x + (2+\varepsilon-k)\hat{y} - \mu x\hat{y} + (k+\mu)\hat{y}^2, \end{array} \quad (63)$$

where $\hat{y} = x + y'$, so $A = -(1+\varepsilon)$, $B = 2+\varepsilon-k$, $g_{11} = -\mu$, $g_{02} = k+\mu$ and $g_{jk} = 0$ otherwise.

We have written a program[12] in the computer algebra language Maple to obtain $\text{Re}(\bar{\lambda} n_{21})$:

```
# stab_index - performs stability index calculation for a Hopf bifurcation.
# We need only go to 3rd order for n21, the term we are after.
q := 3:
# Construct prenormal form
zprime := lambda*z:
zbarprime := lambdabar*zbar:
for j from 0 to q do
        for i from 0 to q do
                if ((i + j >= 2) and (i + j <= q)) then
                        zprime := zprime + a.i.j*z**i*zbar**j:
                        zbarprime := zbarprime + abar.i.j*zbar**i*z**j:
                fi:
        od:
od:
# Transform to normal form coordinates
w := z:
wbar := zbar:
for j from 0 to q do
        for i from 0 to q do
                if ((i + j >= 2) and (i + j <= q) and (i <> j + 1) and
                not((i = 0) and (j = q))) then
                        w := w + p.i.j*z**i*zbar**j:
                        wbar := wbar + pbar.i.j*zbar**i*z**j:
                fi:
        od:
od:
# Construct normal form
coefficients := zprime - lambda*w:
for j from 0 to q do
        for i from 0 to q do
                if ((i + j >= 2) and (i + j <= q)) then
                        if ((i = j + 1) or ((i = 0) and (j = q))) then
                                coefficients := coefficients -
                                        n.i.j*w**i*wbar**j:
                        else
                                coefficients := coefficients +
                                        p.i.j*zprime**i*zbarprime**j:
```

---

[12]This program is now published and made available in the Maple collection.

```
                        fi:
                fi:
        od:
od:
# Coefficients of normal form
readlib(isolate):
coefficients := collect(coefficients,[z,zbar],distributed):
for j from 2 to q do
        for i from 0 to j do
                if ((2*i = j + 1) or ((i = 0) and (j = q))) then
                        assign(isolate(coeff(coeff(coefficients,z,i)
                                ,zbar,j-i),n.i.(j-i))):
                else
                        assign(isolate(coeff(coeff(coefficients,z,i)
                                ,zbar,j-i),p.i.(j-i))):
                        pbar.i.(j-i) := conjugate(p.i.(j-i)):
                fi:
        od:
od:
# Put Bogdanov map into form for calculation:
# x' = y
# y' = A x + B y + g20 x^2 + g11 x y + g02 y^2 + ...
epsilon := 0:                                   # On unit circle
# Linear coefficients
A := -1-epsilon:
B := 2+epsilon-k:
# First set all nonlinear coefficients to zero
for j from 0 to q do
        for i from 0 to q do
                if ((i + j >= 2) and (i + j <= q)) then
                        g.i.j := 0:
                fi:
        od:
od:
# Then explicitly set the couple that are actually nonzero
g11 := -mu:
g02 := k+mu:
# Put the map into complex (prenormal) form
cos(alpha) := B/2:
sin(alpha) := sqrt(-A-(B/2)**2):
lambda := cos(alpha)+I*sin(alpha):               # On unit circle
lambdabar := 1 / lambda:
xprime := y:
yprime := A*x+B*y:
for j from 0 to q do
        for i from 0 to q do
                if ((i + j >= 2) and (i + j <= q)) then
                        yprime := yprime + g.i.j*x**i*y**j:
```

```
                        fi:
                od:
od:
x := (z+zbar)/(2*sin(alpha)):
y := (lambda*z+lambdabar*zbar)/(2*sin(alpha)):
zprime := collect(I*xprime*lambdabar-I*yprime,[z,zbar],distributed):
# Coefficients of prenormal form
for j from 0 to q do
        for i from 0 to q do
                if ((i + j >= 2) and (i + j <= q)) then
                        a.i.j := coeff(coeff(zprime,z,i),zbar,j):
                        abar.i.j := conjugate(a.i.j):
                fi:
        od:
od:
# Print Re_lambdabar_n21; the term that determines stability -
# we have a supercritical Hopf bifurcation if Re_lambdabar_n21 < 0,
# or a subcritical Hopf bifurcation if Re_lambdabar_n21 > 0.
Re_lambdabar_n21 := simplify(evalc(Re(lambdabar*n21)));
```

On running the program, we obtain the answer (which we have also obtained manually):

```
    |\^/|        MAPLE V
._|\|   |/|_.    Copyright (c) 1981-1990 by the University of Waterloo.
 \  MAPLE  /     All rights reserved.  MAPLE is a registered trademark of
 <____ ____>     Waterloo Maple Software.
      |          Type ? for help.
> read stab_index;
                                       mu
              Re_lambdabar_n21 := - 1/2 -----------
                                        k (- 4 + k)
```

Since $0 < k < 4$, we can see that the stability index $\text{Re}(\bar{\lambda} n_{21}) = \mu/(2k(4-k))$ will be positive for $\mu$ positive, and negative for $\mu$ negative, so the Hopf bifurcation is subcritical for $\mu > 0$ and supercritical for $\mu < 0$. The fixed point we obtained in Eq.(61) gives us a Hopf circle in the polar coordinates, which is attracting in the supercritical case and repelling in the subcritical case. In the $(x,y)$ coordinates the circle $w\bar{w} = r^2$ becomes the ellipse $x^2 - 2xy\cos\alpha + y^2 = r^2$.